\documentclass[
reprint,
dcolumn,
bm,
nofootinbib,
notitlepage,
superscriptaddress]{revtex4-2}
\usepackage{graphicx}
\usepackage{color}
\usepackage[latin1]{inputenc}
\usepackage[T1]{fontenc}
\usepackage{amsmath}
\usepackage{amssymb}
\usepackage{braket}
\usepackage{mathrsfs}
\usepackage{hyperref}
\usepackage{ulem}
\usepackage{upgreek}
\usepackage{pstricks-add}
\usepackage{natbib} 
\usepackage{listings}
\usepackage{tikz}
\usepackage{tcolorbox}
\usetikzlibrary{mindmap}

\definecolor{lightblue}{rgb}{0.145,0.6666,1}

\begin{document}
\title{Cavity-modified electronic interactions in molecular ensembles under vibrational strong coupling: Combined insights from cavity Born-Oppenheimer perturbation and \textit{ab initio} wave function theories}

\author{Eric W. Fischer}
\email{ericwfischer.sci@posteo.de}
\affiliation{Institut f\"ur Chemie, Humboldt-Universit\"at zu Berlin, Brook-Taylor-Stra\ss{}e 2, D-12489, Berlin, Germany}

\date{\today}

\let\newpage\relax

\begin{abstract}
Resonant vibrational strong coupling (VSC) between molecular vibrations and quantized field modes of low-frequency optical cavities constitutes the conceptual cornerstone of vibro-polaritonic chemistry. In this work, we theoretically investigate complementary nonresonant electron-cavity mode-interactions in the cavity Born-Oppenheimer (CBO) approximation. We focus on cavity-induced modifications of local and non-local electronic interactions in dipole-coupled molecular ensembles under VSC. Methodologically, we combine CBO perturbation theory (CBO-PT) [Fischer, Saalfrank, \textit{JCTC} \textbf{19}, 7215 (2023)] with non-perturbative CBO Hartree-Fock and coupled cluster theories. Wave function approaches are reformulated to self-consistently address a cavity reaction potential (CRP), which minimizes the electronic energy in the cavity subspace. We derive up to second-order CBO-PT corrections of intra- and intermolecular energies revealing non-trivial corrections to dipole-dipole, dipole-induced-dipole and van-der-Waals interactions, and provide analytical second-order CRP for unimolecular and interacting bimolecular scenarios. In the unimolecular case, we find small local modifications of molecular PES for selected isomerization reactions dominantly captured by the first-order dipole fluctuation. Excellent agreement between CBO-PT and non-perturbative wave function results is obtained indicating minor VSC-induced state relaxation effects in the single-molecule limit. In the bimolecular scenario, CBO-PT reveals an explicit coupling of interacting dimers to cavity modes besides cavity-polarization dependent dipole-induced-dipole and van-der-Waals interactions with enhanced long-range character. An illustrative  CBO-CCSD-based numerical analysis of selected molecular dimer models provides a complementary non-perturbative perspective on cavity-modified intermolecular interactions under VSC. 
\end{abstract}

\let\newpage\relax
\maketitle
\newpage

\section{Introduction}
\label{sec.introduction}
The emerging field of vibro-polaritonic chemistry relies on \textit{resonant} vibrational strong coupling (VSC) between infrared active molecular vibrational modes and confined light modes of an infrared Fabry-P\'erot cavity.\cite{hirai2020,nagarajan2021,dunkelberger2022} The VSC regime is characterized by light-matter hybrid states known as vibrational polaritons\cite{ebbesen2016}, which have been first experimentally characterized by means of vibrational spectroscopies\cite{shalabney2015,george2015,long2015} and moreover observed to intriguingly alter ground state chemistry\cite{thomas2016,thomas2019,ahn2023}. In this work, we theoretically investigate the role of complementary \textit{nonresonant} interactions between low-frequency cavity modes and electrons in vibro-polaritonic chemistry. Specifically, we focus on cavity-induced modifications of electronic interactions under VSC in a Born-Oppenheimer type scenario with respect to both intra- and intermolecular properties by combining perturbative and non-perturbative wave function based approaches.

A theoretical framework, which consistently describes both VSC and electron-photon interactions is provided by the cavity Born-Oppenheimer (CBO) approach.\cite{flick2017,flick2017cbo,fischer2023a} The interacting light-matter many-body system is here decomposed into two coupled subsystems, the high energy (``fast'') electrons and low energy (``slow'') nuclei and cavity modes. The electronic subsystem is described by an CBO electronic Hamiltonian, which accounts for electron-photon interactions, and the complementary nuclear-cavity problem fully accounts for VSC.\cite{flick2017cbo} An adiabatic ground state theory is obtained via the CBO approximation, which straightforwardly generalizes ideas from molecular quantum mechanics to the VSC light-matter hybrid scenario.\cite{flick2017cbo,fischer2023a} In the CBO approximation, electron-photon interactions are naturally nonresonant, since the fundamental electronic excitation energy is by assumption significantly larger than its cavity mode equivalent involved in VSC.

A valid CBO approximation indicates \textit{weak} electron-photon interactions in analogy to weak vibronic coupling in the molecular BO scenario. We recently exploited this observation to formulate a perturbative approach to the CBO electronic ground state problem named CBO perturbation theory (CBO-PT).\cite{fischer2023a,fischer2024} In this work, we discuss CBO-PT corrections of cavity-modified electronic interactions in dipole-coupled molecular ensembles under VSC, \textit{i.e.}, local ground state cavity potential energy surfaces (cPES) and intermolecular interactions, by deriving up to second order analytical expressions. Our results extend recent related studies of the electronic strong coupling (ESC) regime and its low-frequency limit.\cite{philbin2023,haugland2023} 

Methodologically, we combine CBO-PT with \textit{ab initio} wave function theory, specifically, CBO Hartree-Fock (CBO-HF)\cite{schnappinger2023} and CBO coupled cluster theory (CBO-CC)\cite{angelico2023}, which provide a non-perturbative numerical reference of the CBO electronic ground state problem. We reformulate both approaches to self-consistently address a cavity reaction potential (CRP)\cite{fischer2022}, which minimizes the total energy in the cavity mode subspace and follows from a simple but formally exact relation to the ground state cPES. The cavity reaction potential fully captures static cavity-induced modifications of the molecular PES and depends only on the nuclear configuration.

We first address cavity-induced intramolecular modifications of electronic properties under VSC in the single-molecule limit. We provide an analytical expression for a unimolecular CRP up to second-order of CBO-PT, which allows us in combination with non-perturbative CBO wave function approaches to extract the small but locally dominant dipole fluctuation correction. Moreover, this complementary perspective identifies VSC-induced state relaxation effects to be of minor relevance in the single-molecule limit at both mean-field and correlated levels of wave function theory. 

Second, we investigate cavity-induced modifications of intermolecular interactions based on the analytical CRP of an interacting molecular dimer under VSC at second order of CBO-PT. We obtain cavity-induced second-order modifications of dipole-dipole, dipole-induced-dipole and van-der-Waals interactions with an enhanced long-range character partially found also for the ESC scenario\cite{philbin2023,haugland2023}. A complementary numerical analysis of interaction potentials for selected polar and non-polar molecular dimer models based on CRP-type CBO-RHF and CBO-CCSD approaches provides further insight from a non-perturbative perspective. Weak intermolecular two-body interactions are observed to be easier modified due to the presence of low-frequency cavity modes and explicit dipole-type dimer-cavity interactions can provide additional energy transfer channels between molecular and cavity subsystems.

The paper is structured as follows. In Sec.\ref{sec.theory}, we first introduce the CBO framework and CBO-PT followed by a discussion of analytical cavity-modified intra- and intermolecular potentials up to second order of CBO-PT. In Sec.\ref{subsec.crp}, we introduce the cavity reaction potential (CRP) and discuss related self-consistent reformulations of non-perturbative CBO-HF and CBO-CCSD approaches in Sec.\ref{eq.nonpert_wf}. In Secs.\ref{sec.cavity_intramolecular} and\ref{sec.cavity_intermolecular}, we provide detailed discussions of intramolecular and intermolecular interactions under VSC based on a combined analysis employing CBO-PT and \textit{ab initio} wave function approaches for unimolecular and bimolecular CRPs. Finally, Sec.\ref{sec.conclusion} concludes this work.

\section{Perturbation Theory}
\label{sec.theory}

\subsection{The CBO Ground State Problem}
\label{subsec.cbo_ground_state}
We consider a light-matter hybrid system composed of molecules strongly coupled to effective transverse field modes of a low-frequency optical cavity. The corresponding Pauli-Fierz Hamiltonian in length gauge representation and long-wavelength approximation reads\cite{flick2017cbo}
\begin{align}
\hat{H}
&=
\hat{T}_c
+
\hat{T}_n
+
\hat{H}_{ec}
\quad,
\label{eq.cbo_pauli_fierz}
\end{align}
with cavity and nuclear kinetic energy operators (KEO), $\hat{T}_c$ and $\hat{T}_n$. The CBO electronic Hamiltonian, $\hat{H}_{ec}$, enters a time-independent Schr\"odinger equation (TISE)\cite{flick2017cbo}
\begin{align}
\hat{H}_{ec}
\ket{\Psi^{(ec)}_\nu(\underline{R},\underline{x})}
&=
E^{(ec)}_\nu(\underline{R},\underline{x})
\ket{\Psi^{(ec)}_\nu(\underline{R},\underline{x})}
\quad,
\label{eq.electron_photon_tise}
\end{align}
which describes the electronic subsystem in presence of both nuclei and the quantized cavity radiation field. Eq.\eqref{eq.electron_photon_tise} provides access to cavity potential energy surfaces (cPES), $E^{(ec)}_\nu(\underline{R},\underline{x})$, and corresponding CBO adiabatic electronic states, $\ket{\Psi^{(ec)}_\nu(\underline{R},\underline{x})}$, that parametrically depend on both nuclear and cavity displacement coordinates, $\underline{R}$ and $\underline{x}$, respectively. In the following, we restrict the discussion to the ground state cPES, $E^{(ec)}_0$, which provides a potential for the strongly coupled nuclear-cavity subsystem\cite{flick2017cbo,fischer2023a} 
\begin{multline}
\left(
\hat{T}_n
+
\hat{T}_c
+
E^{(ec)}_0(\underline{R},\underline{x})
\right)
\chi^{(nc)}_{i0}(\underline{R},\underline{x})
\\
+
\sum_{\mu>0}
\hat{\mathcal{K}}^{(nc)}_{0\mu}\,
\chi^{(nc)}_{i\mu}(\underline{R},\underline{x})
=
\varepsilon_i\,
\chi^{(nc)}_{i0}(\underline{R},\underline{x})
\quad,
\label{eq.nuclear_cavity_tise}
\end{multline}
with (ro)vibro-polaritonic states, $\chi^{(nc)}_{i0}(\underline{R},\underline{x})$, energies, $\varepsilon_i$, and non-adiabatic coupling elements, $\hat{\mathcal{K}}^{(nc)}_{0\mu}$. In analogy to the Born-Oppenheimer framework in molecular quantum mechanics, TISEs \eqref{eq.electron_photon_tise} and \eqref{eq.nuclear_cavity_tise} describe high energy (``fast'') and low energy (``slow'') degrees of freedom. While resonant VSC between cavity modes and rovibrational degrees of freedom is captured by Eq.\eqref{eq.nuclear_cavity_tise}, the focus of this work lies on Eq.\eqref{eq.electron_photon_tise}, which relates to the \textit{dressed} electronic subsystem interacting with low-frequency cavity modes. We now assume a ground state cPES energetically well separated from the adiabatic excited state manifold
\begin{align}
\Delta^{(ec)}_1
\gg
\Delta^{(nc)}_1
\quad,
\label{eq.low_frequency_condition}
\end{align}
with $\Delta^{(ec)}_1=E^{(ec)}_1-E^{(ec)}_0$ and $\Delta^{(nc)}_1=\varepsilon_1-\varepsilon_0$, which motivates the cavity Born-Oppenheimer approximation\cite{flick2017cbo,fischer2023a} 
\begin{align}
\hat{\mathcal{K}}^{(nc)}_{0\mu}
\propto
(\Delta^{(ec)}_{\mu})^{-1}
\approx
0
\quad,
\quad
\mu\geq 1
\quad.
\label{eq.cbo_approx}
\end{align}
An appealing consequence of the CBO approximation relates to the observation that \textit{nonresonant} interactions between electrons and low-frequency cavity modes are \textit{weak}. This is in close analogy to the molecular BO approximation, which renders respective vibronic interactions weak. Accordingly, we can exploit perturbative theory to investigate cavity-induced modifications of electronic interactions in molecular ensembles under VSC, where we specifically consider the recently formulated CBO perturbation theory (CBO-PT)\cite{fischer2023a}.

\subsection{CBO Perturbation Theory for Dipole-Coupled Molecular Ensembles}
\label{sec.cbopt_ensemble}
In the CBO approximation, the electronic ground state problem for an ensemble of $M$ dipole-coupled molecules interacting with the quantized field of a low-frequency optical cavity is described by an ensemble CBO electronic Hamiltonian
\begin{align}
\hat{H}_{ec}
&=
\hat{H}^{(e)}_\mathrm{ens}
+
V_c
+
\hat{V}_\mathrm{ens}
+
\hat{W}_\mathrm{ens}
\quad.
\label{eq.ensemble_el_ph_hamilton}
\end{align}
The first term simply constitutes the electronic Hamiltonian of the non-interacting molecular ensemble
\begin{align}
\hat{H}^{(e)}_\mathrm{ens}
&=
\sum^M_a
\hat{H}^{(e)}_a
\quad,
\label{eq.ensemble_electron_photon_hamiltonian}
\end{align}
with single-molecule electronic contributions
\begin{align}
\hat{H}^{(e)}_a
&=
\hat{T}^{(e)}_a
+
\hat{V}^{(ee)}_a
+
\hat{V}^{(en)}_a
+
\hat{V}^{(nn)}_a
\quad,
\end{align} 
determined by the electronic KEO, $\hat{T}^{(e)}_a$, besides Coulomb potentials for electron-electron, $\hat{V}^{(ee)}_a$, electron-nuclei, $\hat{V}^{(en)}_a$, and nuclear-nuclear interactions, $\hat{V}^{(nn)}_a$, respectively. The second term in Eq.\eqref{eq.ensemble_el_ph_hamilton} is the harmonic cavity mode potential
\begin{align}
V_c(\underline{x})
&=
\dfrac{1}{2}
\sum^{2N_c}_{\lambda,k}
\omega^2_k\,
x^2_{\lambda k}
\quad,
\end{align}
for $2N_c$ cavity modes characterized by harmonic frequencies, $\omega_k$, and cavity displacement coordinates, $x_{\lambda k}$, with mode index, $k$, and polarization index, $\lambda$. Every cavity mode, $k$, is doubly degenerate with respect to two orthogonal polarization directions, $\lambda=1,2$. The third term in Eq.\eqref{eq.ensemble_el_ph_hamilton} resembles the intermolecular interaction potential
\begin{align}
\hat{V}_\mathrm{ens}
&=
\dfrac{1}{2}
\sum^M_{a,b\neq a}
\hat{V}_{ab}
\quad,
\end{align}
which we approximate by a dipole-dipole interaction\cite{haugland2023}
\begin{align}
\hat{V}_{ab}
&=
-
\sum_{\kappa,\kappa^\prime}
\hat{d}^a_\kappa
\mathcal{T}^{\kappa\kappa^\prime}_{ab}\,
\hat{d}^b_{\kappa^\prime}
\quad,
\label{eq.dipdip_int}
\end{align}
with Cartesian indices, $\kappa,\kappa^\prime$, Cartesian dipole operator components, $\hat{d}^a_\kappa$ and $\hat{d}^b_{\kappa^\prime}$, as well as the intermolecular interaction tensor
\begin{align}
\mathcal{T}^{\kappa\kappa^\prime}_{ab}
&=
-
\dfrac{1}{R^3_{ab}}
\left(
\delta_{\kappa\kappa^\prime}
-
\dfrac{3\,R^\kappa_{ab} R^{\kappa^\prime}_{ab}}{R^2_{ab}}
\right)
\quad.
\label{eq.T_tensor}
\end{align}
The latter is characterized by the intermolecular distance, $R_{ab}=\vert\underline{R}_{ab}\vert$, and a direction vector, $\underline{R}_{ab}$, with Cartesian components, $R^\kappa_{ab}$ and $R^{\kappa^\prime}_{ab}$. The dipole interaction in Eq.\eqref{eq.dipdip_int} captures the long-range components of intermolecular interactions, which we will discuss here with respect to cavity-induced modifications, but does not cover the short-range interaction component left for future studies.

The last term in Eq.\eqref{eq.ensemble_el_ph_hamilton} constitutes the ensemble light-matter interaction potential
\begin{align}
\hat{W}_\mathrm{ens}
&=
\sum^M_a
\hat{W}_a
+
\sum^M_{a,b\neq a}
\hat{W}_{ab}
\quad,
\end{align} 
with one- and two-body contributions
\begin{align}
\hat{W}_a
&=
-
g_0
\sum^{2N_c}_{\lambda,k}
\omega_k\,
\hat{d}^a_{\lambda k}\,
x_{\lambda k}
+
\dfrac{g^2_0}{2}
\sum^{2N_c}_{\lambda,k}
\hat{d}^a_{\lambda k}
\hat{d}^a_{\lambda k}
\quad,
\vspace{0.2cm}
\\
\hat{W}_{ab}
&=
\dfrac{g^2_0}{2}
\sum^{2N_c}_{\lambda,k}
\hat{d}^a_{\lambda k}
\hat{d}^b_{\lambda k}
\quad.
\end{align}
The one-body term, $\hat{W}_a$, contains the light-matter interaction linear in the light-matter interaction constant, $g_0=\frac{1}{\sqrt{\epsilon_0 V_\mathrm{cav}}}$, with cavity volume, $V_\mathrm{cav}$, and the permittivity, $\epsilon_0$, which is determined by a polarization-projected dipole operator, $\hat{d}^a_{\lambda k}=\underline{e}_{\lambda k}\cdot\underline{\hat{d}}^{(en)}_a$, for a cavity mode polarization vector, $\underline{e}_{\lambda k}$, and molecular dipole operator, $\underline{\hat{d}}^{(en)}_a=\underline{\hat{d}}^{(e)}_a+\underline{\hat{d}}^{(n)}_a$. Further, the second term of $\hat{W}_a$ is the intramolecular dipole self-energy (DSE) quadratic in $g_0$, whereas $\hat{W}_{ab}$ resembles an \textit{intermolecular} DSE contribution, which augments the molecular two-body interaction, $\hat{V}_{ab}$, in Eq.\eqref{eq.dipdip_int}. 

In CBO-PT, the CBO electronic Hamiltonian in Eq.\eqref{eq.ensemble_el_ph_hamilton} is partitioned as\cite{fischer2023a}
\begin{align}
\hat{H}_{ec}
&=
\hat{H}_0
+
\lambda\,
\hat{V}_1
\quad,
\end{align}
with perturbation parameter $\lambda$ equals unity, zeroth-order Hamiltonian, $\hat{H}_0$, and perturbation, $\hat{V}_1$, given by
\begin{align}
\hat{H}_0
&=
\hat{H}^{(e)}_\mathrm{ens}
+
V_c
\quad,
\label{eq.cbopt_reference}
\vspace{0.2cm}
\\
\hat{V}_1
&=
\hat{V}_\mathrm{ens}
+
\hat{W}_\mathrm{ens}
\quad.
\label{eq.cbopt_perturbation}
\end{align}
The zeroth-order Hamiltonian, $\hat{H}_0$, is with Eq.\eqref{eq.ensemble_electron_photon_hamiltonian} separable in the ensemble electronic subspace since the harmonic cavity potential, $V_c$, constitutes just an additive constant energy shift. We consider zeroth-order reference states as direct products of single-molecule adiabatic electronic states assuming that individual molecules are sufficiently far apart in agreement with the dipole-dipole interaction. The perturbation, $\hat{V}_1$, contains \textit{both} the ensemble light-matter interaction contributions, $\hat{W}_a$ and $\hat{W}_{ab}$, \textit{and} the molecular dipole-dipole interaction, $\hat{V}_{ab}$, which allows for electron-photon correlation corrections on both intra- and intermolecular electronic interactions.\cite{philbin2023} At $n^\mathrm{th}$-order of CBO-PT, we write CBO-PT($n$) with $n=0,1,2,\dots$

\subsection{Ensemble CBO-PT Energies}
\label{subsec.ensemble_cbopt_energies}
The second order CBO-PT ground state energy for a dipole-coupled molecular ensemble under VSC approximates the corresponding ensemble cPES as
\begin{align}
E^{(ec)}_0
\approx
\mathcal{E}^{(2)}_0
=
\mathcal{E}^{(2)}_\mathrm{intra}
+
\mathcal{E}^{(2)}_\mathrm{inter}
\quad,
\end{align}
where the first term on the right-hand side constitutes the local intramolecular energy contribution 
\begin{align}
\mathcal{E}^{(2)}_\mathrm{intra}
&=
\sum^M_a
\left(
E^{(e)}_a
+
E^{(1)}_a
+
E^{(2)}_a
\right)
+
V_c
\quad,
\label{eq.intra_energies_cbopt}
\end{align}
which will be given explicitly in Sec.\ref{subsubsec.intra_cbopt_energy}. The second term resembles the non-local intermolecular energy contribution, which can be written in analogy to a many-body expansion as
\begin{align}
\mathcal{E}^{(2)}_\mathrm{inter}
&=
\sum^M_{a\neq b}
\left(
E^{(1)}_{ab}
+
E^{(2)}_{ab}
+
U^{(2)}_{ab}
\right)
+
\sum^M_{a\neq b\neq c}
E^{(2)}_{abc}
\quad,
\label{eq.inter_energies_cbopt}
\end{align}
with explicit expressions for all intermolecular interactions being given in Sec.\ref{subsubsec.inter_cbopt_energy}. We note, two- and three-body corrections, $E^{(n)}_{ab}$ and $E^{(2)}_{abc}$, in Eq.\eqref{eq.inter_energies_cbopt} are related to CBO-PT contributions where only excited states of one molecule contribute. In contrast, the pair-interaction energy, $U^{(2)}_{ab}$, is determined by excited states from both molecules. An explicit derivation of all CBO-PT corrections is provided in the supplementary information (SI), Sec.S1.

\subsubsection{Intramolecular CBO-PT Energies}
\label{subsubsec.intra_cbopt_energy}
The intramolecular energy, $\mathcal{E}^{(2)}_\mathrm{intra}$, in Eq.\eqref{eq.intra_energies_cbopt} collects contributions from all $M$ molecules and their interactions with the cavity field. The first-order energy correction, $E^{(1)}_a$, contains the light-matter interaction and the ground-state DSE term 
\begin{align}
E^{(1)}_a
&=
-
g_0
\sum^{2N_c}_{\lambda,k}
\omega_k\,
d^{0_a}_{\lambda k}\,
x_{\lambda k}
+
\dfrac{g^2_0}{2}
\sum^{2N_c}_{\lambda,k}
\mathcal{F}^{0_a}_{\lambda k}
\quad,
\label{eq.cbopt1_energy}
\end{align}
with
\begin{align}
\mathcal{F}^{0_a}_{\lambda k}
&=
d^{0_a}_{\lambda k}
d^{0_a}_{\lambda k}
+
\tilde{\mathcal{F}}^{0_a}_{\lambda k}
\quad.
\label{eq.dse_cbopt1}
\end{align}
Here, $d^{0_a}_{\lambda k}$, is the polarization-projected ground state dipole moment of the $a^\mathrm{th}$-molecule and $\tilde{\mathcal{F}}^{0_a}_{\lambda k}$ is the corresponding dipole fluctuation contribution
\begin{align}
\tilde{\mathcal{F}}^{0_a}_{\lambda k}
&=
\braket{
\Psi^{(e)}_0
\vert
\hat{d}^{(e)}_{\lambda k}
\hat{d}^{(e)}_{\lambda k}
\vert
\Psi^{(e)}_0}
-
\braket{
\Psi^{(e)}_0
\vert
\hat{d}^{(e)}_{\lambda k}
\vert
\Psi^{(e)}_0}^2
\quad,
\label{eq.dipole_fluctuation}
\end{align}
which is purely electronic in character.\cite{haugland2023} The dipole fluctuation term is regularly neglected in a mean-field type approximation underlying effective ground state Hamiltonians for vibro-polaritonic chemistry, which we recently introduced as crude CBO approximation.\cite{fischer2023a} The related crude ground state cPES does not capture cavity-induced changes of classical activation barriers but accounts for VSC-induced distortions leading for example to barrier broadening.\cite{fischer2023a,fischer2021,fischer2022}

The intramolecular second-order CBO-PT energy correction, $E^{(2)}_a$, decomposes into three terms
\begin{align}
E^{(2)}_a
&=
-
\dfrac{g^2_0}{2}
\sum^{2N_c}_{\lambda,k}
\sum^{2N_c}_{\lambda^\prime,k^\prime}
\omega_k\,
x_{\lambda k}\,
\alpha^{0_a}_{\lambda\lambda^\prime}\,
x_{\lambda^\prime k^\prime}\,
\omega_{k^\prime}
\label{eq.cbopt2_intra_intint}
\vspace{0.2cm}
\\
&\hspace{0.4cm}
+
\dfrac{g^3_0}{2}
N_c
\sum^{N_c}_k
\sum^2_{\lambda,\lambda^\prime}
\omega_k\,
x_{\lambda k}\,
\mathcal{A}^{0_a}_{\lambda\lambda^\prime}
\label{eq.cbopt2_intra_intdse}
\vspace{0.2cm}
\\
&\hspace{0.4cm}
-
\dfrac{g^4_0}{8}
N^2_c
\sum^2_{\lambda,\lambda^\prime}
\mathcal{B}^{0_a}_{\lambda\lambda^\prime}
\label{eq.cbopt2_intra_dsedse}
\quad.
\end{align}
The first term is determined by polarization-projected static polarizability tensor elements, $\alpha^{0_a}_{\lambda\lambda^\prime}$, and constitutes a matter-induced screening of harmonic cavity frequencies ($k=k^\prime$) besides a matter-mediated inter-cavity mode coupling ($k\neq k^\prime$).\cite{fischer2023a} The interaction and DSE corrections in Eqs.\eqref{eq.cbopt2_intra_intdse} and \eqref{eq.cbopt2_intra_dsedse} contain linear response functions, $\mathcal{A}^{0_a}_{\lambda\lambda^\prime}$ and $\mathcal{B}^{0_a}_{\lambda\lambda^\prime}$, which can be written as (\textit{cf.} Appendix \ref{subsec.details_cbopt_corr})
\begin{align}
\mathcal{A}^{0_a}_{\lambda\lambda^\prime}
&=
\alpha^{0_a}_{\lambda\lambda^\prime}
d^{0_a}_{\lambda^\prime}
+
\tilde{\mathcal{A}}^{0_a}_{\lambda\lambda^\prime}
\quad,
\label{eq.cbopt2_A}
\vspace{0.2cm}
\\
\mathcal{B}^{0_a}_{\lambda\lambda^\prime}
&=
d^{0_a}_\lambda
\alpha^{0_a}_{\lambda\lambda^\prime}
d^{0_a}_{\lambda^\prime}
+
\tilde{\mathcal{B}}^{0_a}_{\lambda\lambda^\prime}
\quad.
\label{eq.cbopt2_B}
\end{align}
In a mean-field approximation, which we will exploit in Secs.\ref{sec.cavity_intramolecular} and \ref{sec.cavity_intermolecular}, the fluctuation contributions $\tilde{\mathcal{A}}^{0_a}_{\lambda\lambda^\prime}$ and $\tilde{\mathcal{B}}^{0_a}_{\lambda\lambda^\prime}$ are neglected.

\subsubsection{Intermolecular CBO-PT Energies}
\label{subsubsec.inter_cbopt_energy}
We now consider the intermolecular CBO-PT(2) energy contribution, $\mathcal{E}^{(2)}_\mathrm{inter}$, in Eq.\eqref{eq.inter_energies_cbopt}, where the first-order correction is given by
\begin{align}
E^{(1)}_{ab}
&=
\dfrac{1}{2}
V_{ab}
+
W_{ab}
\quad,
\end{align}
with 
\begin{align}
V_{ab}
&=
-
\sum_{\kappa,\kappa^\prime}
d^{0_a}_\kappa
\mathcal{T}^{\kappa\kappa^\prime}_{ab}\,
d^{0_b}_{\kappa^\prime}
\quad,
\vspace{0.2cm}
\\
W_{ab}
&=
\dfrac{g^2_0}{2}
\sum^{2N_c}_{\lambda,k}
d^{0_a}_{\lambda k}
d^{0_b}_{\lambda k}
\quad,
\end{align}
which relate to a distance-dependent dipole-dipole interaction, $V_{ab}$, besides a distance-\textit{independent} DSE-induced bimolecular interaction, $W_{ab}$. The second-order two-body contribution, $E^{(2)}_{ab}$, contains two distinct terms
\begin{align}
E^{(2)}_{ab}
&=
\tilde{E}^{(2)}_{ab}
+
\tilde{E}^{(2)}_{abb}
\quad,
\label{eq.cbopt2_two_body}
\end{align}
where, $\tilde{E}^{(2)}_{ab}$, decomposes into
\begin{align}
\tilde{E}^{(2)}_{ab}
&=
-
g_0
\sum^{2N_c}_{\lambda,k}
\omega_k\,
x_{\lambda k}\,
\sum_{\kappa,\kappa^\prime}
\alpha^{0_a}_{\lambda\kappa}\,
\mathcal{T}^{\kappa\kappa^\prime}_{ab}
d^{0_b}_{\kappa^\prime}
\label{eq.cbopt2_Eab_interaction_g1}
\vspace{0.2cm}
\\
&\hspace{0.4cm}
+
\dfrac{g^2_0}{2}
N_c
\sum^2_{\lambda}
\sum_{\kappa,\kappa^\prime}
\mathcal{A}^{0_a}_{\lambda\kappa}
\mathcal{T}^{\kappa\kappa^\prime}_{ab}
d^{0_b}_{\kappa^\prime}
\label{eq.cbopt2_Eab_interaction_g2}
\vspace{0.2cm}
\\
&\hspace{0.4cm}
+
g^3_0\,
N_c
\sum^{2N_c}_{\lambda,k}
\sum^{2}_{\lambda^\prime}
\omega_k\,
x_{\lambda k}\,
\alpha^{0_a}_{\lambda\lambda^\prime}\,
d^{0_b}_{\lambda^\prime}
\label{eq.cbopt2_Eab_interaction_g3}
\vspace{0.2cm}
\\
&\hspace{0.4cm}
-
\dfrac{g^4_0}{2}
N^2_c
\sum^2_{\lambda,\lambda^\prime}
\mathcal{A}^{0_a}_{\lambda\lambda^\prime}
d^{0_b}_{\lambda^\prime}
\quad.
\label{eq.cbopt2_Eab_interaction_g4}
\end{align}
Here, the terms linear and quadratic in $g_0$ dependent via $\mathcal{T}^{\kappa\kappa^\prime}_{ab}$ on the intermolecular distance, $R_{ab}$, in contrast to distance-independent contributions cubic and quartic in $g_0$. Moreover, terms scaling as $g_0$ and $g^3_0$ explicitly couple interacting molecular dimers via the cavity displacement coordinate, $x_{\lambda k}$, to the cavity field. The two-body three-index term, $\tilde{E}^{(2)}_{abb}$, in Eq.\eqref{eq.cbopt2_two_body} follows as special case of the three-body correction 
\begin{align}
E^{(2)}_{abc}
&=
-
\dfrac{1}{2}
\sum_{\kappa,\kappa^\prime}
\sum_{\gamma,\gamma^\prime}
\mathcal{T}^{\kappa\kappa^\prime}_{ab}\,
\mathcal{T}^{\gamma\gamma^\prime}_{ac}
d^{0_b}_{\kappa^\prime}\,
\alpha^{0_a}_{\kappa\gamma}\,
d^{0_c}_{\gamma^\prime}
\label{eq.three_body_1}
\vspace{0.2cm}
\\
&\hspace{0.4cm}
+
g^2_0\,
N_c
\sum^2_{\lambda}
\sum_{\kappa,\kappa^\prime}
\mathcal{T}^{\kappa\kappa^\prime}_{ab}
d^{0_b}_{\kappa^\prime}
\alpha^{0_a}_{\kappa\lambda}\,
d^{0_c}_{\lambda}
\label{eq.three_body_2}
\vspace{0.2cm}
\\
&\hspace{0.4cm}
-
\dfrac{g^4_0}{2}
N^2_c
\sum^2_{\lambda,\lambda^\prime}
d^{0_b}_{\lambda}
\alpha^{0_a}_{\lambda\lambda^\prime}
d^{0_c}_{\lambda^\prime}
\label{eq.three_body_3}
\quad,
\end{align}
with $b=c$, and can be interpreted as cavity-modified dipole-induced-dipole interaction (\textit{cf.} Sec.\ref{sec.cavity_intermolecular}). Finally, the second-order pair interaction correction reads
\begin{align}
U^{(2)}_{ab}
&=
U^{ab}_\mathrm{vdW}
+
U^{ab}_\mathrm{cvdW}
+
U^{ab}_\mathrm{cav}
\quad,
\end{align}
and contains three individual terms
\begin{align}
U^{ab}_\mathrm{vdW}
&=
-
\dfrac{1}{2}
\sum_{\kappa,\kappa^\prime}
\sum_{\gamma,\gamma^\prime}
\mathcal{T}^{\kappa\kappa^\prime}_{ab}
\mathcal{T}^{\gamma\gamma^\prime}_{ab}
\alpha^{ab}_{\kappa\gamma\kappa^\prime\gamma^\prime}
\quad,
\vspace{0.2cm}
\\
U^{ab}_\mathrm{cvdW}
&=
g^2_0
\sum^2_\lambda
\sum_{\kappa,\kappa^\prime}
\mathcal{T}^{\kappa\kappa^\prime}_{ab}
\alpha^{ab}_{\kappa\lambda\kappa^\prime\lambda}
\quad,
\vspace{0.2cm}
\\
U^{ab}_\mathrm{cav}
&=
-
\dfrac{g^4_0}{2}
N^2_c
\sum^2_{\lambda,\lambda^\prime}
\alpha^{ab}_{\lambda\lambda^\prime\lambda\lambda^\prime}
\quad.
\end{align}
The first term resembles the common van-der-Waals (vdW) interaction, $U^{ab}_\mathrm{vdW}$, which is augmented by a cavity-induced long-range vdW-type component, $U^{ab}_\mathrm{cvdW}$, and a purely cavity-related distance independent term $U^{ab}_\mathrm{cav}$. Similar corrections for cavity-modified van-der-Waals interactions have been derived recently for the ESC regime.\cite{philbin2023,haugland2023}

With intra- and intermolecular second-order CBO-PT energies at hand, we will now discuss cavity-induced modifications of electronic properties in molecules. For this purpose, we first introduce the notion of a cavity reaction potential, which provides a representation of the cPES beneficial for the discussion of chemical reactivity under VSC.

\section{Cavity Reaction Potentials}
\label{subsec.crp}
We write the non-perturbative ground state cPES, $E^{(ec)}_0$, in Eq.\eqref{eq.electron_photon_tise} as
\begin{align}
E^{(ec)}_0(\underline{R},\underline{x})
&=
\mathcal{V}^{(ec)}_\mathrm{crp}(\underline{R})
+
\dfrac{1}{2}
\Delta \underline{x}\,
\underline{\underline{H}}^{(ec)}_{CC}(\underline{R})
\Delta \underline{x}
\quad,
\label{eq.taylor_cpes}
\end{align}
which resembles a second-order Taylor expansion around a cavity reference coordinate
\begin{align}
x^{(ec)}_0
&=
\dfrac{g_0}{\omega_k}
\braket{
\Psi^{(ec)}_0
\vert
\hat{d}_\lambda
\vert
\Psi^{(ec)}_0
}
\quad,
\label{eq.crp_min_cav_coordinate}
\end{align}
subject to two equivalent constraints\cite{schnappinger2023}
\begin{align}
\left.
\dfrac{\partial E^{(ec)}_0}{\partial x_{\lambda k}}
\right\vert_{\underline{x}^{(ec)}_0}
=
0
\quad
\Leftrightarrow
\quad
\braket{
\Psi^{(ec)}_0
\vert
\underline{\hat{E}}_\perp
\vert
\Psi^{(ec)}_0
}
=
0
\quad.
\label{eq.crp_min_cav_conditions_exact}
\end{align} 
The first constraint minimizes the energy with respect to variations along all cavity displacement coordinates and eliminates the linear first-order term in Eq.\eqref{eq.taylor_cpes}. The second constraint ensures a ``non-radiating ground state'', \textit{i.e.}, it imposes a vanishing ground state expectation value of the transverse electrical component, $\underline{\hat{E}}_\perp$, of the cavity field.\cite{schaefer2020,schnappinger2023} The relation in Eq.\eqref{eq.taylor_cpes} in combination with constraints in Eq.\eqref{eq.crp_min_cav_conditions_exact} is formally exact, since the CBO electronic Hamiltonian is at most quadratic in cavity displacement coordinates such that all higher-order terms of the Taylor expansion vanish identically. 

The first term in Eq.\eqref{eq.taylor_cpes} is denoted as cavity reaction potential (CRP)\cite{fischer2022} 
\begin{align}
\mathcal{V}^{(ec)}_\mathrm{crp}(\underline{R})
&=
E^{(ec)}_0(\underline{R},\underline{x}^{(ec)}_0)
\quad,
\label{eq.crp}
\end{align}
which accounts for \textit{static} cavity-induced modifications of the molecular potential energy surface while minimizing the CBO electronic energy with respect to variations along cavity displacement coordinates. Due to the harmonic nature of the cavity potential, we have here always an energetic minimum with respect to the cavity coordinates. The notion of \textit{static} cavity-induced modifications relates to effects that do no involve cavity mode excitations, which would require a cavity coordinate dependence. Therefore, $\mathcal{V}^{(ec)}_\mathrm{crp}$ allows for addressing cavity-induced modifications of chemical reactions in terms of changes in nuclear configuration space. 

The second term in Eq.\eqref{eq.taylor_cpes} contains the cavity component of the CBO Hessian, $\underline{\underline{H}}^{(ec)}_{CC}$\cite{bonini2022,fischer2024} with coordinate vector, $\Delta \underline{x}=\underline{x}-\underline{x}^{(ec)}_0$. This term captures the full potential around the CRP in molecular and cavity coordinate space and accordingly describes excitations in the cavity mode subspace, which underlie cavity-induced modifications of \textit{dynamic} processes.\cite{li2021a,fischer2022,sun2022,schaefer2022a,fischer2023b,lindoy2023,ying2023,ke2024} We like to point out that the notion of static/dynamic refers here to time-independent/time-dependent processes and does not relate to static/dynamic correlation as used in the context of electronic structure theory.

The non-perturbative CRP in Eq.\eqref{eq.crp} can be explicitly written as
\begin{align}
\mathcal{V}^{(ec)}_\mathrm{crp}(\underline{R})
&=
\braket{
\Psi^{(ec)}_0
\vert
\hat{H}_e
\vert
\Psi^{(ec)}_0}
+
\dfrac{g^2_0}{2}
\mathcal{\tilde{F}}^{(ec)}_\lambda
\quad,
\label{eq.exact_crp}
\end{align}
where the second term corresponds to a non-perturbative dipole fluctuation contribution 
\begin{align}
\mathcal{\tilde{F}}^{(ec)}_\lambda
&=
\braket{
\Psi^{(ec)}_0
\vert
\hat{d}_\lambda
\hat{d}_\lambda
\vert
\Psi^{(ec)}_0}
-
\braket{
\Psi^{(ec)}_0
\vert
\hat{d}_\lambda
\vert
\Psi^{(ec)}_0}^2
\quad.
\end{align}
Moreover, the cavity-induced correction of the molecular PES is simply given by
\begin{align}
\Delta^{(ec)}_\mathrm{crp}(\underline{R};\lambda)
&=
\mathcal{V}^{(ec)}_\mathrm{crp}(\underline{R};\lambda)
-
E^{(e)}_0(\underline{R})
\quad,
\label{eq.nonperturb_energy_shift_bare}
\end{align} 
which vanishes in the non-interacting limit, $g_0\to0$. In Secs.\ref{sec.cavity_intramolecular} and \ref{sec.cavity_intermolecular}, we will discuss approximations to the non-perturbative CRP in Eq.\eqref{eq.exact_crp} from two complementary perspectives: First, a CBO-PT(2) approach based on the related second-order cPES derived in Sec.\ref{subsec.ensemble_cbopt_energies}. In this context, we employ a CBO-PT approximation of constraints in Eq.\eqref{eq.crp_min_cav_conditions_exact}, which relates the non-radiating ground state condition to a mean-field approximation of Eq.\eqref{eq.cbopt2_A}. Second, as non-perturbative reference, we consider slight reformulations of recently introduced \textit{ab initio} CBO wave function approaches\cite{schnappinger2023,angelico2023}, which here self-consistently account for constraints in Eq.\eqref{eq.crp_min_cav_conditions_exact} and directly solve the CBO electronic TISE for the CRP, $\mathcal{V}^{(ec)}_\mathrm{crp}$.

\section{Ab initio CBO Wave Function Theory for Cavity Reaction Potentials}
\label{eq.nonpert_wf}
We introduce reformulations of CBO Hartree-Fock (CBO-HF) theory\cite{schnappinger2023} and CBO coupled cluster theory with singles and doubles (CBO-CCSD)\cite{angelico2023}, which provide a numerical route to the non-perturbative CRP, $\mathcal{V}^{(ec)}_\mathrm{crp}$. Both approaches have been implemented based on the PySCF software package\cite{sun2018,sun2020}. 

\subsection{CBO Hartree-Fock Theory for CRPs}
\label{subsec.crp_cbo_hf}
We reformulate CBO-HF theory by self-consistently including the minimizing cavity coordinate in Eq.\eqref{eq.crp_min_cav_conditions_exact} in the CBO-Fockian, which directly provides access to a mean-field approximation of the CRP in Eq.\eqref{eq.exact_crp}. We consider a single Slater-Determinant, $\Psi^{(ec)}_\mathrm{cbohf}$, which allows us to obtain the CBO-RHF CRP as
\begin{align}
\mathcal{V}^{(ec)}_\mathrm{crp}(\underline{R})
&=
\underset{\underline{x},\underline{\kappa}}{\mathrm{min}}
\braket{
\Psi^{(ec)}_\mathrm{cbohf}(\underline{\kappa})
\vert
\hat{H}_{ec}(\underline{x})
\vert
\Psi^{(ec)}_\mathrm{cbohf}(\underline{\kappa})
}
\quad.
\label{eq.min_cbohf}
\end{align}
The energy expectation value is here simultaneously minimized with respect to both orbital rotation parameters, $\underline{\kappa}$, and cavity-displacement coordinates, $\underline{x}$. We restrict the following discussion to the restricted CBO-HF scenario. From Eq.\eqref{eq.min_cbohf}, we obtain a restricted CRP-CBO Fockian 
\begin{align}
\tilde{f}^{(e)}_\lambda
&=
\sum_{pq}
\left(
\tilde{h}^{pq}_\lambda
+
\tilde{v}^{pq}_\lambda
\right)
\hat{E}_{pq}
\quad,
\end{align}
where we restrict the discussion to a single cavity mode with polarization index, $\lambda$. A generalization to multi-cavity-mode systems as discussed in Sec.\ref{sec.theory} is straightforward. Core and mean-field potential contributions for a closed-shell system are given by
\begin{align}
\tilde{h}^{pq}_\lambda
&=
h_{pq}
+
\dfrac{g^2_0}{2}
O^{pq}_\lambda
\quad,
\vspace{0.2cm}
\\
\tilde{v}^{pq}_\lambda
&=
J_{pq}
-
\dfrac{1}{2}
\tilde{K}^{pq}_\lambda
\quad.
\end{align}
Here, the bare electronic core contribution, $h_{pq}$, is augmented by the 1-particle contribution of the DSE, $O^{pq}_\lambda$ (\textit{cf.} Appendix \ref{subsec.ab_initio_cbopt1}) and the mean-field potential contains the bare Coulomb contribution, $J_{pq}$, besides a DSE-corrected exchange term 
\begin{align}
\tilde{K}^{pq}_\lambda
=
\sum_{rs}
D_{rs}
\left(
g_{prsq}
+
g^2_0\,
d^{pr}_\lambda
d^{sq}_\lambda
\right)
\quad,
\end{align}
with one-particle reduced density matrix, $D_{rs}$. The CRP CBO-RHF ground state energy reads (\textit{cf.} Appendix \ref{subsec.ab_initio_cbopt1})
\begin{align}
\mathcal{V}^\mathrm{cbohf}_\mathrm{crp}(\underline{R})
&=
E^{(e)}_\mathrm{cbohf}
+
g^2_0
\left(
\sum_i
O^{ii}_\lambda
-
\sum_{ij}
d^{ij}_\lambda
d^{ji}_\lambda
\right)
\,,
\label{eq.cbo_rhf_energy}
\end{align}
which is independent of both cavity mode frequencies and cavity displacement coordinates. Here, $E^{(e)}_\mathrm{cbohf}$ is the RHF energy evaluated with respect to CBO-RHF molecular orbitals, which account for DSE-induced relaxation effects. The second and third terms account for local and exchange-type DSE contributions. Moreover, the corresponding mean-field approximation of the cavity-induced energy correction is given by $\Delta^\mathrm{cbohf}_\mathrm{crp}=\mathcal{V}^\mathrm{cbohf}_\mathrm{crp}-E^{(e)}_0$.

\subsection{CBO Coupled Cluster Theory for CRPs}
The CBO coupled cluster approach with singles and doubles corrections (CBO-CCSD)\cite{angelico2023} is determined by the wave function ansatz
\begin{align}
\ket{\Psi^{(ec)}_\mathrm{cbocc}}
&=
\exp\left(\hat{T}_1+\hat{T}_2\right)
\ket{\Psi^{(ec)}_\mathrm{cbohf}}
\quad,
\label{eq.cbo_ccsd_ansatz}
\end{align}
and allows to account for electron correlation under VSC. Here, $\hat{T}_1$ and $\hat{T}_2$ are the common singles and doubles excitation operators but $\Psi^{(ec)}_\mathrm{cbohf}$ is the CRP-CBO-HF reference state. In spirit of the CRP-CBO-HF approach, we consider a slight reformulation of CBO-CCSD to obtain the non-perturbative CRP as
\begin{align}
\mathcal{V}^\mathrm{cbocc}_\mathrm{crp}(\underline{R})
&=
\underset{\underline{x}}{\mathrm{min}}
\braket{
\Psi^{(ec)}_\mathrm{cbocc}
\vert
\hat{H}_{ec}
\vert
\Psi^{(ec)}_\mathrm{cbocc}}
\quad,
\end{align}
which leads to an effective CBO electronic Hamiltonian 
\begin{align}
\hat{H}^0_{ec}
&=
\sum_{pq}
\tilde{h}^0_{pq}
\hat{E}_{pq}
+
\dfrac{1}{2}
\sum_{pqrs}
\tilde{g}_{pqrs}
\hat{e}_{pqrs}
+
\tilde{V}_0
\quad,
\label{eq.crp_cbo_hamiltonian}
\end{align}
where indices $p,q,r,s$ run over the molecular orbital basis and the constant energy shift reads
\begin{align}
\tilde{V}_0
&=
V_{nn}
+
\dfrac{g^2_0}{2}
\braket{
\Psi^{(ec)}_0
\vert
\hat{d}_\lambda
\vert
\Psi^{(ec)}_0
}^2
\quad,
\end{align}
with nuclear Coulomb interaction potential, $V_{nn}$. Augmented 1- and 2-electron integrals are given by
\begin{align}
\tilde{h}^0_{pq}
&=
\hat{h}_{pq}
+
\dfrac{g^2_0}{2}
O^{pq}_\lambda
-
g^2_0\,
d^{pq}_\lambda
\braket{
\Psi^{(ec)}_0
\vert
\hat{d}_\lambda
\vert
\Psi^{(ec)}_0
}
\quad,
\end{align}
and
\begin{align}
\tilde{g}_{pqrs}
&=
g_{pqrs}
-
g^2_0\,
d^{pq}_\lambda
d^{rs}_\lambda
\quad.
\end{align}
Since $\hat{H}^0_{ec}$ depends on the adiabatic ground state solution via the dipole expectation value, the CRP CBO-CCSD approach is solved self-consistently starting from a dipole moment guess obtained from the underlying CRP CBO-RHF calculation. A detailed discussion will be subject of a future study.

\section{Cavity-Modified Intramolecular Interactions}
\label{sec.cavity_intramolecular}
We apply CBO-PT and CRP-type reformulations of CBO wave function approaches to address cavity-induced modifications of intramolecular electronic interactions for selected single molecule examples. Specifically, we analyse VSC-induced energy changes of molecular PES in terms of a unimolecular CRP based on both CBO-PT(2) and non-perturbative CBO-RHF/CBO-CCSD approaches in CRP-type formulation. 
\begin{figure*}[hbt]
\begin{center}
\includegraphics[scale=1.0]{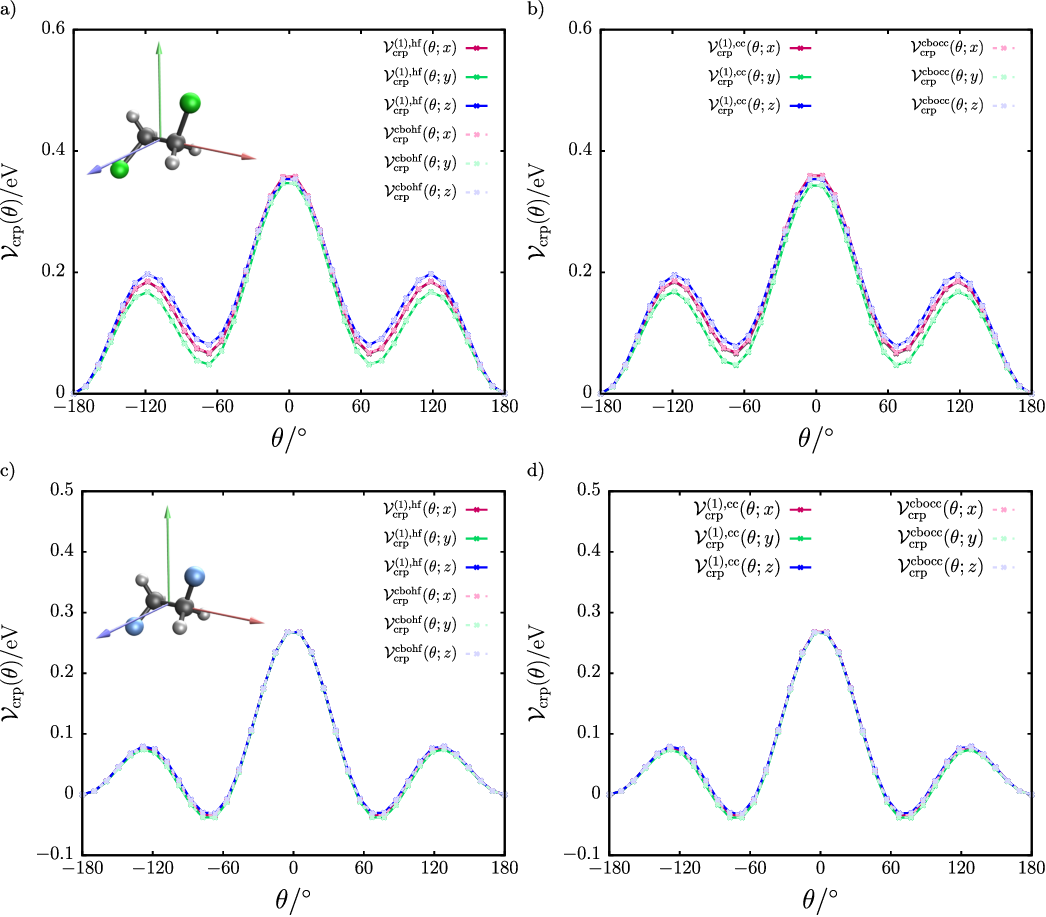}
\end{center}
\renewcommand{\baselinestretch}{1.}
\caption{Torsion PES $E^{(e)}_0(\theta)$ obained via TPSSh/cc-pVTZ in black and cavity-polarization dependent CRPs of (a,b) 1,2-DCE (carbon in black, hydrogen in white, chlorine in green) and (c,d) 1,2-DFE (fluorine in blue) under VSC evaluated via (a,c) CBO-PT(1)-RHF/cc-pVDZ and CBO-RHF/cc-pVDZ and (b,d) CBO-PT(1)-CCSD/cc-pVDZ and CBO-CCSD/cc-pVDZ with light-matter coupling strength, $g_0=0.03\sqrt{E_h}/e\,a_0$, for cavity-mode polarizations $\lambda=x$ (red), $\lambda=y$ (green) and $\lambda=z$ (blue). The CRP is independent of the cavity mode frequency.}
\label{fig.torsion_cpes}
\end{figure*}

\subsection{Unimolecular Cavity Reaction Potential}
\label{subsec.uni_crp}
We first derive the unimolecular CRP at second-order of CBO-PT with details provided in the SI, Sec.S2. We consider a single molecule coupled to a single cavity mode with coordinate $x_c$ and polarization $\lambda$. The corresponding CBO-PT(2) cPES reads
\begin{align}
\mathcal{E}^{(2)}_0
&=
E^{(e)}_0
+
V_c
+
E^{(1)}_0
+
E^{(2)}_0
\quad,
\label{eq.cbopt_2_monomer}
\end{align}
where first- and second-order CBO-PT energy corrections are explicitly given by
\begin{align}
E^{(1)}_0
&=
-
g_0\,
\omega_c\,
d^0_\lambda
x_c
+
\dfrac{g^2_0}{2}
d^0_\lambda
d^0_\lambda
+
\dfrac{g^2_0}{2}
\tilde{\mathcal{F}}^0_\lambda
\quad,
\label{eq.monomer_cbopt1}
\end{align}
and 
\begin{align}
E^{(2)}_0
&=
-
\dfrac{g^2_0}{2}
\omega^2_c\,
\alpha^0_{\lambda\lambda}\,
x^2_c
+
\dfrac{g^3_0}{2}
\omega_c\,
x_c\,
\mathcal{A}^0_{\lambda\lambda}
-
\dfrac{g^4_0}{8}
\mathcal{B}^0_{\lambda\lambda}
\quad.
\label{eq.monomer_cbopt2}
\end{align}
The second-order minimizing cavity coordinate, $x^{(2)}_0$, for the unimolecular problem is obtained as (\textit{cf.} SI, Secs.S2.A,B)
\begin{align}
x^{(2)}_0
&=
\dfrac{g_0}
{\omega_c}
d^0_\lambda
+
\dfrac{g^3_0}
{2\omega_c}
\alpha^0_{\lambda \lambda}
d^0_\lambda
\quad,
\label{eq.unimolecular_cbopt2_mincav}
\end{align}
where the CBO-PT(2) non-radiating ground state condition imposes a mean-field approximation on the linear response function, $\mathcal{A}^0_{\lambda\lambda}$, in Eq.\eqref{eq.cbopt2_A}. The unimolecular CBO-PT(2) cavity reaction potential is then derived from Eqs.\eqref{eq.cbopt_2_monomer} and \eqref{eq.unimolecular_cbopt2_mincav} as (\textit{cf.} SI, Sec.S2.C) 
\begin{align}
\mathcal{V}^{(2)}_\mathrm{crp}(\underline{R};\lambda)
=
E^{(e)}_0(\underline{R})
+
\dfrac{g^2_0}{2}
\tilde{\mathcal{F}}^0_\lambda(\underline{R})
-
\dfrac{g^4_0}{8}
\mathcal{B}^0_{\lambda\lambda}(\underline{R})
\quad,
\label{eq.unimolecular_cbopt2_crp}
\end{align}
where we made the parametric dependence of $\mathcal{V}^{(2)}_\mathrm{crp}$ on the cavity-mode polarization index, $\lambda$, explicit. The first term on the right-hand side is simply the molecular ground state PES, $E^{(e)}_0$, which is in leading-order corrected by the first-order dipole fluctuation, $\tilde{\mathcal{F}}^0_\lambda$. The third term on the right-hand side reflects a contribution to the second-order CBO-PT correction determined by the linear response function, $\mathcal{B}^0_{\lambda\lambda}$ (\textit{cf.} Appendix \ref{subsec.details_cbopt_corr}).

\subsection{Isomerization under VSC} 
\label{subsec.num_res_iso}
Illustratively, we analyse unimolecular CRPs for the isomerization of 1,2-Dichloroethane (1,2-DCE) and its less polarizable fluorinated equivalent 1,2-Difluoroethane (1,2-DFE) under VSC with $g_0=0.03\sqrt{E_h}/e\,a_0$ following recent work\cite{fischer2024}. Computational details are provided in Appendix \ref{subsec.compute_details}. We will compare cavity-induced energy corrections of the molecular PES obtained from CBO-PT and non-perturbative wave function approaches, which allows us to address the quality of the dipole fluctuation correction and the relevance of higher-order CBO-PT effects in the single-molecule limit.

An appealing connection between CBO-PT and the non-perturbative approach can be established by reformulating the non-perturbative energy correction in Eq.\eqref{eq.nonperturb_energy_shift_bare} as (\textit{cf.} Appendix \ref{subsec.nonperturb_energy_shift})
\begin{align}
\Delta^{(ec)}_\mathrm{crp}(\underline{R};\lambda)
&=
\dfrac{g^2_0}{2}
\mathcal{\tilde{F}}^0_\lambda(\underline{R})
+
\Delta^{(ec)}_\mathrm{rlx}(\underline{R};\lambda)
\quad,
\label{eq.nonperturb_energy_shift}
\end{align}
where the right-hand side contains here simply the CBO-PT(1) dipole fluctuation correction as given in Eq.\eqref{eq.dipole_fluctuation} and a second term
\begin{align}
\Delta^{(ec)}_\mathrm{rlx}(\underline{R};\lambda)
&=
\mathcal{V}^{(ec)}_\mathrm{crp}(\underline{R};\lambda)
-
\mathcal{V}^{(1)}_\mathrm{crp}(\underline{R};\lambda)
\quad.
\label{eq.relax_correction}
\end{align}
From Eqs.\eqref{eq.exact_crp} and \eqref{eq.unimolecular_cbopt2_crp} (up to the second-order correction), we realize that both expressions on the right-hand side differ only in the type of adiabatic ground state, \textit{i.e.}, CBO adiabatic vs. bare adiabatic, such that $\Delta^{(ec)}_\mathrm{rlx}$ can be interpreted as energy correction capturing cavity-induced state relaxation effects. 

We now compare cavity-induced corrections of the torsion PES/CRP on both CBO-RHF/CBO-PT(1)-RHF and CBO-CCSD/CBO-PT(1)-CCSD levels of theory with a focus on the first-order dipole-fluctuation correction. The color code in Fig.\ref{fig.torsion_cpes} refers to the single-cavity mode polarization chosen parallel to respective axis of the molecular coordinate system. Motivated by recent theoretical results pointing at the potential relevance of cavity-altered structural relaxation and reorientation of molecules under strong coupling\cite{liebenthal2024,schnappinger2024,lexander2024}, we explicitly consider the polarization-dependence of the dipole fluctuation correction instead of addressing rotationally averaged CRPs.

For both examples, we find an excellent agreement between CBO-PT(1) and \textit{ab initio} wave function theory at both mean-field and correlated levels of theory, which indicates that cavity-induced intramolecular energy corrections are captured at first-order of CBO-PT in the single-molecule limit. Moreover, we find only small numerical differences with respect to non-perturbative results, which indicates minor state relaxation effects, \textit{i.e.}, minor contributions of $\Delta^{(ec)}_\mathrm{rlx}$. From a chemical perspective, we find small but clearly observable polarization-dependent corrections for the torsion PES of the easier polarizeable 1,2-DCE system (\textit{cf.} Fig.\ref{fig.torsion_cpes}a,b) in contrast to 1,2-DFE, where cavity-induced energy modifications are significantly weaker and hardly observable from Fig.\ref{fig.torsion_cpes}c,d. 

Previous perturbative results for 1,2-DCE under VSC\cite{galego2019} formally relied on a dipole-mediated second-order energy correction of the torsion PES, which is traced back to a neglected DSE term in the Hamiltonian, whose relevance has been discussed in\cite{rokaj2018,schaefer2020}. This finding is to be contrasted by the dipole fluctuation correction, which is the dominant local correction in our analysis, and can be in particular non-zero for molecules without permanent dipole moment. The role of collective effects manifesting in altered local changes of electronic properties under VSC, as recently indicated by mean-field studies\cite{schnappinger2023,sidler2024}, and their potentially non-perturbative character are subject of a future project.

\section{Cavity-Modified Intermolecular Interactions}
\label{sec.cavity_intermolecular}
We now turn to cavity-induced modifications of intermolecular interactions under VSC, which are captured only at second-order of CBO-PT in contrast to the dominant first-order dipole fluctuation correction discussed in Sec.\ref{sec.cavity_intramolecular}.

\subsection{Interacting Dimer under VSC}
\label{subsec.bi_crp}
We consider an interacting dimer composed of molecules A and B at distance $R$ under VSC with a single cavity mode. The corresponding CBO-PT(2) cPES reads
\begin{align}
\mathcal{E}^{(2)}_\mathrm{dimer}
&=
\tilde{\mathcal{E}}^{(2)}_A
+
\tilde{\mathcal{E}}^{(2)}_B
+
V_c
+
\mathcal{E}^{(2)}_{AB}
\quad,
\label{eq.cbopt_2_dimer}
\end{align}
with unimolecular contributions, $\tilde{\mathcal{E}}^{(2)}_A$ and $\tilde{\mathcal{E}}^{(2)}_B$, as given in Eq.\eqref{eq.cbopt_2_monomer} excluding the cavity potential, $V_c$. The intermolecular interaction is written as
\begin{align}
\mathcal{E}^{(2)}_{AB}
&=
V^{(2)}_{AB}
+
\tilde{V}^{(2)}_{AB}
+
U^{(2)}_{AB}
+
W^{(2)}_{AB}
\quad,
\label{eq.cbopt_2_dimer_interaction}
\end{align}
and contains four distinct contributions, which we consider along a fixed Cartesian axis, $\kappa$, such that the intermolecular interaction tensor in Eq.\eqref{eq.T_tensor} turns into $\mathcal{T}^{\kappa\kappa}_{AB}=2R^{-3}$. The first term, $V^{(2)}_{AB}$, relates to a cavity-altered dipole-dipole interaction
\begin{align}
V^{(2)}_{AB}(R)
&=
-
\dfrac{D_3}{R^3}
-
g_0
\dfrac{D^\prime_{3\lambda}\,\omega_c}{R^3}
x_c
+
g^2_0
\dfrac{D^{\prime\prime}_{3\lambda}}{R^3}
\quad,
\label{eq.vab_int}
\end{align}
where the second and third term follow from Eqs.\eqref{eq.cbopt2_Eab_interaction_g1} and \eqref{eq.cbopt2_Eab_interaction_g1}, and coefficients are explicitly given in Appendix \ref{subsec.coeff_cbopt2_intpot}. The first correction ($D^\prime_3$) is linear in the light-matter coupling constant, $g_0$, and constitutes a cavity-mediated dipole-polarizability type interaction. This term directly couples the molecular dimer via the coordinate, $x_c$, to the cavity field and therefore provides an energy transfer channel, which can directly influence the dynamics of the cavity subsystem. The second correction ($D^{\prime\prime}_3$) resembles a higher-order interaction between two polar molecules, which additionally involves polarizability effects.

The second interaction term in Eq.\eqref{eq.cbopt_2_dimer_interaction} is distance-independent and exclusively contains cavity-induced contributions (\textit{cf.} Appendix \ref{subsec.coeff_cbopt2_intpot})
\begin{align}
\tilde{V}^{(2)}_{AB}
&=
g^2_0\,
\tilde{D}_{3\lambda}
+
g^3_0\,
\tilde{D}^\prime_{3\lambda}\,
\omega_c\,
x_c
-
\dfrac{g^4_0}{2}
\tilde{D}^{\prime\prime}_{3\lambda}
\quad,
\label{eq.eab_int}
\end{align}
which are closely related to their counterparts in Eq.\eqref{eq.vab_int}. The first term ($\tilde{D}_3$) constitutes the first-order DSE-induced intermolecular interaction, whereas the second and third term scaling as $g^3_0$ and $g^4_0$ relate to Eqs.\eqref{eq.cbopt2_Eab_interaction_g3} and \eqref{eq.cbopt2_Eab_interaction_g4}. All coefficients entering Eq.\eqref{eq.eab_int} as given in Appendix \ref{subsec.coeff_cbopt2_intpot} are fully determined by the cavity-mode polarization index, $\lambda$, in contrast to the Cartesian axis determining coefficients in Eq.\eqref{eq.vab_int}. 

The third interaction term in Eq.\eqref{eq.cbopt_2_dimer_interaction} relates to a cavity-modified van-der-Waals interaction with long-range character (\textit{cf.} Appendix \ref{subsec.coeff_cbopt2_intpot})
\begin{align}
U^{(2)}_{AB}(R)
&=
-
\dfrac{C_6}{R^6}
+
g^2_0
\dfrac{C_{3\lambda}}{R^3}
-
g^4_0
\dfrac{C_{0\lambda}}{R^0}
\quad,
\label{eq.cvdW_int}
\end{align}
which has been recently derived by Koch and coworkers, however, for the ESC regime.\cite{philbin2023,haugland2023} Here, a long-range correction scaling as $g^2_0\,R^{-3}$ enters besides a \textit{weak} distance-\textit{independent} contribution scaling as $g^4_0$, which can be traced back to the dipole approximation. Finally, the forth contribution in Eq.\eqref{eq.cbopt_2_dimer_interaction} constitutes a cavity-modified dipole-induced dipole interaction (\textit{cf.} Appendix \ref{subsec.coeff_cbopt2_intpot})
\begin{align}
W^{(2)}_{AB}(R)
&=
-
\dfrac{\tilde{C}_6}{R^6}
+
g^2_0
\dfrac{\tilde{C}_{3\lambda}}{R^3}
-
g^4_0
\dfrac{\tilde{C}_{0\lambda}}{R^0}
\quad,
\label{eq.dip_ind_dip}
\end{align}
which results from the two-body contributions in $E^{(2)}_{abb}$ (\textit{cf.} Eqs.\eqref{eq.three_body_1} to \eqref{eq.three_body_3}). At this point, we like to note that there is only a single first order contribution to the cavity-modified intermolecular interaction energy, which is the distance-independent DSE-interaction term in Eq.\eqref{eq.eab_int} with coefficient $\tilde{D}_{3\lambda}$. Thus, in contrast to intramolecular corrections, second-order CBO-PT turns out to be crucial for the intermolecular interaction scenario.
\begin{figure*}[hbt!]
\begin{center}
\includegraphics[scale=1.0]{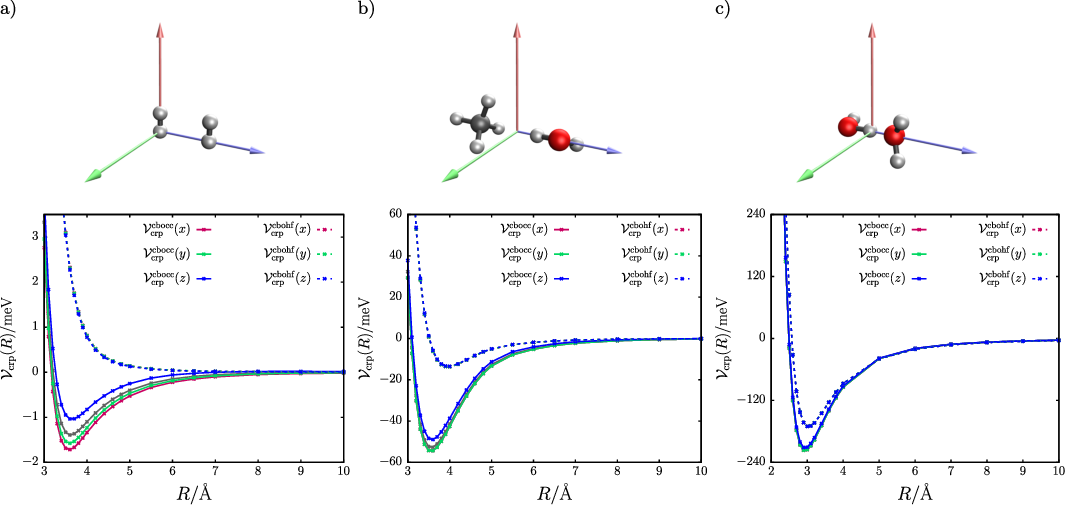}
\end{center}
\renewcommand{\baselinestretch}{1.}
\caption{Interaction CRPs for (a) H$_2$-dimer, (b) CH$_4$-H$_2$O-dimer and (c) H$_2$O-dimer (carbon in black, hydrogen in white, oxygen in red) evaluated via CBO-CCSD/aug-cc-pVDZ at light-matter coupling strength, $g_0=0.03\sqrt{E_h}/e\,a_0$, for cavity-mode polarizations $\lambda=x$ (red), $\lambda=y$ (green) and $\lambda=z$ (blue). Interaction CRPs are obtained by subtracting energies for monomers separated by $50$ \AA and are independent of cavity frequency.}
\label{fig.dissociation_cpes}
\end{figure*}

\subsection{Bimolecular Cavity Reaction Potential}
\label{subsec.bi_crp}
In analogy to the unimolecular scenario in Sec.\ref{subsec.uni_crp}, we will now discuss the cavity reaction potential for the interacting dimer with a focus on cavity-modified intermolecular interactions. Details on the derivation are provided in SI, Sec.S3. The minimizing cavity coordinate obtained for $\mathcal{E}^{(2)}_\mathrm{dimer}$ in Eq.\eqref{eq.cbopt_2_dimer} is given by
\begin{align}
x^{(2)}_0
&=
x^{(2)}_\mathrm{loc}
+
x^{(2)}_\mathrm{int}
\quad,
\label{eq.dimer_cbopt2_mincav}
\end{align}
with a local intramolecular term
\begin{align}
x^{(2)}_\mathrm{loc}
&=
\dfrac{g_0}{\omega_c}
\left(
d^{0_A}_\lambda
+
d^{0_B}_\lambda
\right)
\label{eq.dimer_cbopt2_mincav_loc}
\vspace{0.2cm}
\\
&\hspace{1cm}
+
\dfrac{g^3_0}{2\omega_c}
\left(
\alpha^{0_A}_{\lambda \lambda}
d^{0_A}_\lambda
+
\alpha^{0_B}_{\lambda \lambda}
d^{0_B}_\lambda
\right)
\quad,
\nonumber
\end{align}
and an intermolecular, distance-dependent contribution
\begin{align}
x^{(2)}_\mathrm{int}
&=
\dfrac{g_0\,D^\prime_3}{\omega_c}
\dfrac{1}{R^3}
+
\dfrac{
g^3_0\,
D^\prime_3
\left(
\alpha^{0_A}_{\lambda \lambda}
+
\alpha^{0_B}_{\lambda \lambda}
\right)
}{\omega_c}
\dfrac{1}{R^3}
\quad.
\label{eq.dimer_cbopt2_mincav_int}
\end{align}
Those two expressions provide access to a cavity reaction potential accurate up to terms scaling as $g^4_0$ in line with the second-order cPES in Eq.\eqref{eq.cbopt_2_dimer}. In the dilute-gas limit, the interaction contribution in Eq.\eqref{eq.dimer_cbopt2_mincav_int} vanishes, \textit{i.e.}, $x^{(2)}_\mathrm{int}\to0$ as $R\to\infty$, which leads with Eq.\eqref{eq.cbopt_2_dimer} to the corresponding bimolecular CRP
\begin{align}
\mathcal{V}^{(2)}_\mathrm{crp}
&=
\sum_{i=A,B}
\left(
E^{(e)}_{0_i}
+
\dfrac{g^2_0}{2}
\tilde{\mathcal{F}}^{0_i}_\lambda
-
\dfrac{g^4_0}{8}
\mathcal{B}^{0_i}_{\lambda\lambda}
\right)
+
\mathcal{\tilde{E}}^{(2)}_{AB}
\,.
\end{align}
Here, the summation runs simply over individual contributions of monomers A and B in line with the unimolecular result in Eq.\eqref{eq.unimolecular_cbopt2_crp}. In addition, we find here an interaction term
\begin{align}
\mathcal{\tilde{E}}^{(2)}_{AB}
&=
-
g^4_0
\left(
C_0
+
\tilde{C}_0
\right)
\quad,
\end{align}
which contains both the distance-\textit{independent} van-der-Waals-type correction in Eq.\eqref{eq.cvdW_int} besides the related dipole-induced dipole correction in Eq.\eqref{eq.dip_ind_dip}.

In presence of distance-dependent intermolecular interactions, $\mathcal{\tilde{E}}^{(2)}_{AB}$ turns into
\begin{align}
\mathcal{\tilde{E}}^{(2)}_{AB}
&=
\mathcal{V}^{(2)}_{AB}(R)
+
U^{(2)}_{AB}(R)
+
W^{(2)}_{AB}(R)
\quad,
\end{align}
which contains an additional CRP term
\begin{multline}
\mathcal{V}^{(2)}_{AB}(R)
=
-
\dfrac{D_3
}{R^3}
-
g^2_0
\left(
\dfrac{V_{3\lambda}}{R^3}
+
\dfrac{V_{6\lambda}}{R^6}
\right)
\\
-
\dfrac{g^4_0}{2}
\left(
\dfrac{V^\prime_{3\lambda}}{R^3}
+
\dfrac{V^\prime_{6\lambda}}{R^6}
\right)
\quad,
\label{eq.crp_vab_int}
\end{multline}
besides the full cavity-modified van-der-Waals and dipole-induced-dipole interactions with coefficients being explicit given in Appendix \ref{subsec.coeff_cbopt2_intpot}. The leading-order correction ($g^2_0$) in $\mathcal{V}^{(2)}_{AB}$ contains interactions, which scale with distance as both $R^{-3}$ and $R^{-6}$ in contrast to the second-order cPES, where $R^{-6}$-contributions are independent of light-matter coupling. We now recall that $\mathcal{\tilde{E}}^{(2)}_{AB}$ describes cavity-modified intermolecular interactions under the constraint that the ground state energy is minimized with respect to variations along cavity coordinates, \textit{i.e.}, it captures static modifications of intermolecular interaction potentials. In the following, we provide a numerical analysis of two-body interactions in selected molecular dimers to illustrate the nature of such VSC-induced modifications.

\subsection{Interaction CRPs under VSC}
Illustratively, we consider a H$_2$-dimer (van-der-Waals interaction) and a H$_2$O-dimer (dipole-dipole interaction) under VSC following earlier studies of the ESC regime\cite{angelico2023,haugland2021} besides a H$_2$O-CH$_4$-dimer (dipole-induced-dipole interaction) under VSC based on a molecular example studied in Ref.\cite{metz2019}. Monomers structures are fixed at the equilibrium geometry of the dimers for all intermolecular distances following Ref.\cite{haugland2021}. Computational details are provided in Appendix \ref{subsec.compute_details}.

In Fig.\ref{fig.dissociation_cpes}, we present numerical results obtained via CBO-RHF/aug-cc-pVDZ and CBO-CCSD/aug-cc-pVDZ for interaction CRPs of all three dimers under VSC. Based on a mean-field description, we do not observe any significant cavity-induced modification of interaction CRPs, such that we concentrate in the following on CBO-CCSD results. VSC-induced effects are most pronounced for the low-energy van-der-Waals interaction, since the interaction energy scale increases from van-der-Waals (Fig.\ref{fig.dissociation_cpes}a) over dipole-induced dipole (Fig.\ref{fig.dissociation_cpes}b) to dipole-dipole interactions (Fig.\ref{fig.dissociation_cpes}c). In Figs.\ref{fig.dissociation_cpes}a and \ref{fig.dissociation_cpes}b, a cavity-induced destabilization of the interaction CRPs is found for cavity mode polarizations parallel to the dissociation direction and a stabilizing effect for orthogonal polarization in agreement with Refs.\cite{angelico2023,haugland2021}. A similar trend is found for the H$_2$O-CH$_4$ dimer, whereas VSC-induced changes of the dipole-dipole interaction potential appear to be of minor relevance due to the larger interaction energies. We now recall, that the CRP-type \textit{ab initio} approach provides access to static cavity-induced modifications of interaction potentials, which is reflected by corrected van-der-Waals and dipole-induced dipole interactions in Eqs.\eqref{eq.cvdW_int} and \eqref{eq.dip_ind_dip} and the dipole-dipole type CRP expression in Eq.\eqref{eq.crp_vab_int}. However, although the latter seems to be less sensitive to VSC-induced static effects due to the larger interaction energy scale, the cavity-modified dipole-dipole correction in Eq.\eqref{eq.vab_int} contains a coupling to the cavity subsystem, which indicates its potential relevance for dynamics of the light-matter hybrid system beyond CRPs.

We close this section by discussing our results in the context of recent findings obtained in the ESC regime\cite{haugland2021,philbin2023,haugland2023}. The cavity-modified van-der-Waals interaction in Eq.\eqref{eq.cvdW_int} is traced back to a perturbative second-order correction emerging from the dipole self-energy and the molecular dipole-dipole interaction. Since cavity modes are not explicitly contributing to this interaction, this expression is equivalent for \textit{ab initio} cavity QED and CBO frameworks employed to study ESC and VSC regimes, respectively. This is presumably no longer the case for interaction contributions involving an explicit coupling to the cavity subsystem, which can be understood as follows: In \textit{ab initio} cavity QED\cite{haugland2021}, cavity modes are grouped with electrons as fast degrees of freedom, which leads to a mixed fermionic-bosonic TISE fully accounting for excitations in both electronic and cavity mode subspaces. This is to be contrasted by the CBO framework as illustrated in Sec.\ref{subsec.cbo_ground_state}, where Eq.\eqref{eq.electron_photon_tise} is purely electronic in character and captures cavity-induced static effects on the electronic subsystem whereas excitations of the low-frequency cavity subsystem are described by Eq.\eqref{eq.nuclear_cavity_tise}. 

A connection between CBO and \textit{ab initio} cavity QED approaches has been recently proposed by taking the low-frequency limit of the latter.\cite{haugland2023} We argue that this does not necessarily lead to formally equivalent descriptions when one consistently takes the low-frequency limit \textit{before} invoking a Born-Oppenheimer type approximation. In this case, the mixed electron-cavity TISE in \textit{ab initio} cQED is solved by electron-polariton states with a fundamental excitation energy similar to the cavity mode frequency as recently illustrated in a numerical study\cite{fischer2023a}. Therefore, a BO-type approximation might no longer be valid since non-adiabatic coupling elements can become significant. In addition, there is a conceptual difference in the description of ground state dynamics in the CBO framework and the low-frequency limit of \textit{ab initio} cavity QED. In the former, the vibro-polaritonic subsystem propagates adiabatically on the ground state cPES, whereas vibro-polaritonic time-evolution in \textit{ab initio} cavity QED is inherently non-adiabatic in nature since cavity mode excitations are accounted for in the electron-polariton problem and therefore involve excited adiabatic states by construction. 

\section{Conclusion}
\label{sec.conclusion}
We theoretically examined the role of nonresonant electron-photon interactions in dipole-coupled molecular ensembles under vibrational strong coupling (VSC) for a cavity Born-Oppenheimer (CBO) ground state scenario. Methodologically, we combined CBO perturbation theory (CBO-PT) with \textit{ab initio} CBO wave function theory, specifically, CBO Hartree-Fock (CBO-HF) and coupled cluster (CBO-CCSD) theories, which we reformulated to directly obtain a ground state cavity reaction potential (CRP). We introduced the CRP by a formally exact second-order Taylor expansion of the ground state cavity potential energy surface (cPES) around a cavity reference coordinate, which minimizes energy in the cavity mode subspace and respects the non-radiating ground state condition. This approach allowed us to provide complementary perspectives on VSC-induced modifications of intra- and intermolecular electronic interactions based on both perturbative and non-perturbative results. We derived explicit expressions for cavity-induced correction of intra- and intermolecular interactions in dipole-coupled molecular ensembles under VSC up to second-order CBO-PT, which we subsequently employed to obtain analytical CRPs for unimolecular and interacting bimolecular examples. 

In the single-molecule limit, we find local electronic properties to be dominantly corrected by the first-order CBO-PT dipole fluctuation. An excellent agreement between CBO-PT and non-perturbative mean-field (CBO-HF) and correlated (CBO-CCSD) approaches allowed us to characterize higher-order CBO-PT corrections as encoding cavity-induced state relaxation effects, \textit{i.e.}, changes in molecular orbital coefficients or cluster amplitudes in the present study. For investigated torsion potentials of chlorinated and fluorinated ethane derivatives, we observe small cavity-induced modifications of classical activation barriers, which are more pronounced for the easier polarizable chlorinated compound, besides minor state relaxation effects under VSC. 

In contrast to small first-order modifications of local electronic properties in the single-molecule limit, we obtain cavity-altered dipole-dipole, dipole-induced-dipole and van-der-Waals interactions at second order of CBO-PT. Dipole-dipole interactions are perturbatively identified to contain an explicit coupling to the cavity subsystem, which indicates potentially altered energy transfer processes of an interacting molecular dimer under VSC. Moreover, we obtain dipole-induced dipole and van-der-Waals interactions with enhanced long-range character scaling as $R^{-3}$ and $R^0$, \textit{i.e.}, the latter being distance-independent under the dipole approximation. We complemented our analytical CBO-PT results via a numerical non-perturbative analysis of selected dimer interaction potentials based on CBO-RHF and CBO-CCSD in CRP-formulation. At the mean field level, we do not observe any significant cavity-induced modifications of intermolecular interactions in contrast to CBO-CCSD, where cavity-induced effects are observed to be potentially relevant for weak interactions of van-der-Waals and dipole-induced-dipole type.

The presented analysis provides insight into the potentially non-trivial role of electron-photon interactions in the VSC regime, whose physical manifestation were addressed by means of CBO-PT. It furthermore provides a starting point for addressing collective VSC with respect to cavity-modified intra- and intermolecular electronic interactions from an \textit{ab initio} perspective. In particular, potentially local manifestations of collective VSC effects in the spirit of very recent related investigations\cite{schnappinger2023,sidler2024,castagnola2024} and their role for vibro-polaritonic chemistry will be subject of future work.

\section*{Supplementary Information}
We provide detailed derivations for (i) the second-order CBO-PT energy corrections for dipole-dipole coupled molecular ensembles under VSC, (ii) the unimolecular CRP  and (iii) the bimolecular CRP. 

\section*{Acknowledgements}
E.W. Fischer acknowledges valuable discussions with Michael Roemelt and funding by the Deutsche Forschungsgemeinschaft (DFG, German Research Foundation) through DFG project 536826332.

\section*{Data Availability Statement}
The data that support the findings of this study are available from the corresponding author upon reasonable request.

\section*{Conflict of Interest}
The author has no conflicts to disclose.

\renewcommand{\thesection}{}
\section*{Appendix}

\setcounter{equation}{0}
\renewcommand{\theequation}{\thesubsection.\arabic{equation}}
\subsection{CBO-PT(2) Linear Response Functions and Mean-Field Approximation}
\label{subsec.details_cbopt_corr}
Linear response functions in Eqs.\eqref{eq.cbopt2_intra_intdse} and \eqref{eq.cbopt2_intra_dsedse} are explicitly given by
\begin{align}
\mathcal{A}^0_{\lambda\lambda^\prime}
&=
-
2
\sum_{\mu\neq0}
\dfrac{
\braket{
\Psi^{(e)}_0
\vert
\hat{d}_\lambda
\hat{d}_\lambda
\vert
\Psi^{(e)}_\mu}
\braket{
\Psi^{(e)}_\mu
\vert
\hat{d}_{\lambda^\prime}
\vert
\Psi^{(e)}_0}
}{\Delta^{(e)}_{\mu}}
\quad,
\vspace{0.2cm}
\\
\mathcal{B}^0_{\lambda\lambda^\prime}
&=
-
2
\sum_{\mu\neq0}
\dfrac{
\braket{
\Psi^{(e)}_0
\vert
\hat{d}_{\lambda}
\hat{d}_{\lambda}
\vert
\Psi^{(e)}_\mu}
\braket{
\Psi^{(e)}_\mu
\vert
\hat{d}_{\lambda^\prime}
\hat{d}_{\lambda^\prime}
\vert
\Psi^{(e)}_0}
}{\Delta^{(e)}_{\mu}}
\quad,
\end{align}
and Eqs.\eqref{eq.cbopt2_A} and \eqref{eq.cbopt2_B} are obtained by inserting a resolution of the identity in the adiabatic electronic subspace between products of polarization-projected dipole operators. This leads to the following expressions for fluctuation contributions
\begin{align}
\tilde{\mathcal{A}}^0_{\lambda\lambda^\prime}
&=
-
2
\sum_{\mu,\nu\neq0}
\dfrac{
d^{0\nu}_\lambda
d^{\nu\mu}_\lambda
d^{\mu0}_\lambda
}{\Delta^{(e)}_{\mu}}
\quad,
\vspace{0.2cm}
\\
&=
\mathcal{A}^0_{\lambda\lambda^\prime}
-
\alpha^0_{\lambda\lambda^\prime}
d^0_{\lambda^\prime}
\quad,
\vspace{0.2cm}
\\
\tilde{\mathcal{B}}^0_{\lambda\lambda^\prime}
&=
-
2
\sum_{\mu,\nu,\nu^\prime\neq0}
\dfrac{
d^{0\nu}_\lambda
d^{\nu\mu}_\lambda
d^{\mu\nu^\prime}_\lambda
d^{\nu^\prime0}_\lambda
}{\Delta^{(e)}_{\mu}}
\quad,
\vspace{0.2cm}
\\
&=
\mathcal{B}^0_{\lambda\lambda^\prime}
-
d^0_{\lambda}
\alpha^0_{\lambda\lambda^\prime}
d^0_{\lambda^\prime}
\quad,
\end{align}
where we introduced the notation, $d^{\nu\mu}_\lambda=\braket{\Psi_\nu\vert\hat{d}_\lambda\vert\Psi_\mu}$, for transition dipole matrix elements. The notion of fluctuation terms, which resemble contributions from excited adiabatic states is inspired by the analogy to the dipole fluctuation term, $\tilde{\mathcal{F}}^0_{\lambda k}$, in Eq.\eqref{eq.dse_cbopt1}. Mean-field approximations are obtained by neglecting the fluctuation terms
\begin{align}
\mathcal{A}^0_{\lambda\lambda^\prime}
&\approx
\alpha^0_{\lambda\lambda^\prime}
d^0_{\lambda^\prime}
\,
\Leftrightarrow
\tilde{\mathcal{A}}^0_{\lambda\lambda^\prime}
\approx
0
\label{eq.cbopt2_A_approx}
\vspace{0.2cm}
\\
\mathcal{B}^0_{\lambda\lambda^\prime}
&\approx
d^0_\lambda
\alpha^0_{\lambda\lambda^\prime}
d^0_{\lambda^\prime}
\,
\Leftrightarrow
\tilde{\mathcal{B}}^0_{\lambda\lambda^\prime}
\approx
0
\quad,
\label{eq.cbopt2_B_approx}
\end{align}
where Eq.\eqref{eq.cbopt2_A_approx} is found to be a consequence of the CBO-PT(2) approximate non-radiating ground state condition (\textit{cf.} SI, Secs.S2.B and S3.C).

\setcounter{equation}{0}
\renewcommand{\theequation}{\thesubsection.\arabic{equation}}
\subsection{Ab initio Dipole-Fluctuation Correction}
\label{subsec.ab_initio_cbopt1}
The dipole-fluctuation correction, $\tilde{\mathcal{F}}^0_\lambda$, contains ground state expectation values of 1- and 2-electron terms, which read
\begin{align}
\braket{
0^{(e)}
\vert
\hat{d}^{(e)}_\lambda
\hat{d}^{(e)}_\lambda 
\vert
0^{(e)}}
&=
\braket{
0^{(e)}
\vert
\sum^{N_e}_i
r^2_{i\lambda}
\vert
0^{(e)}}
\vspace{0.2cm}
\\
&\hspace{0.5cm}
+
\braket{
0^{(e)}
\vert
\sum^{N_e}_{i\neq j}
r_{i\lambda}
r_{j\lambda}
\vert
0^{(e)}}
\quad,
\end{align} 
with polarization-projected electronic coordinates, $r_{i\lambda}=\underline{e}_\lambda\cdot\underline{r}_i$. In second quantization, the expectation value turns into 
\begin{align}
\braket{
0^{(e)}
\vert
\hat{d}^{(e)}_\lambda
\hat{d}^{(e)}_\lambda 
\vert
0^{(e)}}
&=
\sum_{pq}
O^{pq}_\lambda
D_{pq}
\vspace{0.2cm}
\\
&\hspace{0.5cm}
+
\sum_{pqrs}
d^{pq}_\lambda	
d^{rs}_\lambda
d_{pqrs}
\quad,
\end{align}
where $p,q,r,s$ run over the molecular orbital basis and 1-electron integrals read
\begin{align}
d^{pq}_\lambda
=
-
\braket{
\psi_p
\vert
r_{1\lambda}
\vert
\psi_q
}
\quad,
\quad
O^{pq}_\lambda
=
\braket{
\psi_p
\vert
r^2_{1\lambda}
\vert
\psi_q}
\quad.
\end{align}
The related expression for the dipole fluctuation correction reads
\begin{align}
\tilde{\mathcal{F}}^0_\lambda
&=
\sum_{pq}
O^{pq}_\lambda
D_{pq}
+
\sum_{pqrs}
d^{pq}_\lambda	
d^{rs}_\lambda
\left(
d_{pqrs}
-
D_{pq}
D_{rs}
\right)
\,,
\end{align}
where individual terms simplify in case of restricted Hartree-Fock, where 1- and 2-particle reduced density matrices, $D_{pq}$ and $d_{pqrs}$, are given by
\begin{align}
D_{pq}
&=
2\delta_{pq}
\quad,
\vspace{0.2cm}
\\
d_{pqrs}
&=
D_{pq}
D_{rs}
-
\dfrac{1}{2}
D_{ps}
D_{rq}
\quad,
\end{align}
to
\begin{align}
\sum_{pq}
O^{pq}_\lambda
D_{pq}
&=
2
\sum_p
O^{pp}_\lambda
\quad,
\vspace{0.2cm}
\\
\sum_{pqrs}
d^{pq}_\lambda	
d^{rs}_\lambda
d_{pqrs}
&=
4
\sum_{pr}
d^{pp}_\lambda	
d^{rr}_\lambda
-
2
\sum_{pr}
d^{pr}_\lambda	
d^{rp}_\lambda
\quad,
\vspace{0.2cm}
\\
\sum_{pq}
d^{pq}_\lambda
D_{pq}
&=
2
\sum_p
d^{pp}_\lambda
\quad.
\end{align}
Thus, the dipole fluctuation correction turns into
\begin{multline}
\tilde{\mathcal{F}}^0_\lambda
=
2
\sum_p
O^{pp}_\lambda
+
4
\sum_{pr}
d^{pp}_\lambda	
d^{rr}_\lambda
-
2
\sum_{pr}
d^{pr}_\lambda	
d^{rp}_\lambda
\\
-
4
\sum_{pr}
d^{pp}_\lambda	
d^{rr}_\lambda
\quad,
\end{multline}
where the forth term stemming from the product of dipole expectation values cancels the second term, which finally gives
\begin{align}
\tilde{\mathcal{F}}^\mathrm{rhf}_\lambda
&=
2
\sum_i
O^{ii}_\lambda
-
2
\sum_{ij}
d^{ij}_\lambda
d^{ji}_\lambda
\quad,
\label{eq.rhf_dip_fluc}
\end{align}
with summation being restricted to occupied MOs.

\setcounter{equation}{0}
\renewcommand{\theequation}{\thesubsection.\arabic{equation}}
\subsection{Computational Details}
\label{subsec.compute_details}

\subsubsection{Monomers}
Chemically accurate torsion potentials for both 1,2-DCE and 1,2-DFE were calculated via TPSSh/cc-pVTZ as implemented in ORCA 5.0.4\cite{orca2012,orcav52022}. Cavity-induced energy corrections based on first-order CBO-PT were calculated via RHF/cc-pVDZ and CCSD/cc-pVDZ as implemented in the PySCF software package\cite{sun2018,sun2020}. Non-perturbative cavity-induced energy corrections were obtained from PySCF-based implementations via CBO-RHF/cc-pVDZ and CBO-CCSD/cc-pVDZ in CRP formulation.

\subsubsection{Dimers}
Monomer structures for the H$_2$-dimer, the H$_2$O-dimer and the H$_2$O-CH$_4$-dimer were fixed at their equilibrium values for all monomer distances following the procedure in Ref.\cite{haugland2021}. For the H$_2$O-dimer, we employed the equilibrium structure from Ref.\cite{ronca2014}. For the H$_2$O-CH$_4$-dimer, we employed the equilibrium structure from Ref.\cite{metz2019}. All dimer interaction potentials were obtained from PySCF-based implementations via CBO-RHF/aug-cc-pVDZ and CBO-CCSD/aug-cc-pVDZ in CRP formulation.

\setcounter{equation}{0}
\renewcommand{\theequation}{\thesubsection.\arabic{equation}}
\subsection{Derivation of Eq.\eqref{eq.nonperturb_energy_shift}}
\label{subsec.nonperturb_energy_shift}
We start from Eq.\eqref{eq.nonperturb_energy_shift_bare}
\begin{align}
\Delta^{(ec)}_\mathrm{crp}
&=
\mathcal{V}^{(ec)}_\mathrm{crp}
-
E^{(e)}_0
\quad,
\vspace{0.2cm}
\\
&=
\dfrac{g^2_0}{2}
\tilde{\mathcal{F}}^0_\lambda
+
\mathcal{V}^{(ec)}_\mathrm{crp}
-
E^{(e)}_0
-
\dfrac{g^2_0}{2}
\tilde{\mathcal{F}}^0_\lambda
\quad,
\vspace{0.2cm}
\\
&=
\dfrac{g^2_0}{2}
\tilde{\mathcal{F}}^0_\lambda
+
\mathcal{V}^{(ec)}_\mathrm{crp}
-
\mathcal{V}^{(1)}_\mathrm{crp}
\quad,
\vspace{0.2cm}
\\
&=
\dfrac{g^2_0}{2}
\tilde{\mathcal{F}}^0_\lambda
+
\Delta^{(ec)}_\mathrm{rlx}
\quad,
\end{align} 
where we added and subtracted the CBO-PT(1) dipole fluctuation correction in the second line. The third line follows from Eq.\eqref{eq.unimolecular_cbopt2_crp} by neglecting the second-order correction
\begin{align}
\mathcal{V}^{(1)}_\mathrm{crp}
&=
E^{(e)}_0
+
\dfrac{g^2_0}{2}
\tilde{\mathcal{F}}^0_\lambda
\quad,
\end{align}
and in the last line, we introduced Eq.\eqref{eq.relax_correction}.

\setcounter{equation}{0}
\renewcommand{\theequation}{\thesubsection.\arabic{equation}}
\subsection{Coefficients for CBO-PT(2) Interaction CRPs}
\label{subsec.coeff_cbopt2_intpot}
Coefficients of the CBO-PT(2) dipole-dipole interaction in Eq.\eqref{eq.vab_int} are explicitly given by
\begin{align}
D_3
&=
2\,
d^{0_A}_\kappa
d^{0_B}_\kappa
\quad,
\vspace{0.2cm}
\\
D^\prime_{3\lambda}
&=
2
\left(
\alpha^{0_A}_{\lambda\kappa}
d^{0_B}_\kappa
+
\alpha^{0_B}_{\lambda\kappa}
d^{0_A}_\kappa
\right)
\quad,
\vspace{0.2cm}
\\
D^{\prime\prime}_{3\lambda}
&=
\alpha^{0_A}_{\lambda\kappa}
d^{0_A}_\kappa
d^{0_B}_\kappa
+
\alpha^{0_B}_{\lambda\kappa}
d^{0_B}_\kappa
d^{0_A}_\kappa
\quad,
\end{align}
where we assume the mean-field approximation in Eq.\eqref{eq.cbopt2_A_approx} for $D^{\prime\prime}_3$. Coefficients of the cavity-induced two-body interaction in Eq.\eqref{eq.eab_int} are given by
\begin{align}
\tilde{D}_{3\lambda}
&=
d^{0_A}_\lambda
d^{0_B}_\lambda
\quad,
\vspace{0.2cm}
\\
\tilde{D}^\prime_{3\lambda}
&=
\alpha^{0_A}_{\lambda\lambda}
d^{0_B}_{\lambda}
+
\alpha^{0_B}_{\lambda\lambda}
d^{0_A}_{\lambda}
\quad,
\vspace{0.2cm}
\\
\tilde{D}^{\prime\prime}_{3\lambda}
&=
\alpha^{0_A}_{\lambda\lambda}
d^{0_A}_\lambda
d^{0_B}_\lambda
+
\alpha^{0_B}_{\lambda\lambda}
d^{0_B}_\lambda
d^{0_A}_\lambda
\quad.
\end{align}
The cavity-modified van-der-Waals interaction in Eq.\eqref{eq.cvdW_int} is determined by coefficients
\begin{align}
C_6
&=
2\alpha^{AB}_{\kappa\kappa\kappa\kappa}
\quad,
\vspace{0.2cm}
\\
C_{3\lambda}
&=
2
\alpha^{AB}_{\kappa\lambda\kappa\lambda}
\quad,
\vspace{0.2cm}
\\
C_{0\lambda}
&=
\dfrac{\alpha^{AB}_{\lambda\lambda\lambda\lambda}
}{2}
\quad,
\end{align}
and the cavity-modified dipole-induced-dipole interaction in Eq.\eqref{eq.dip_ind_dip} contains
\begin{align}
\tilde{C}_6
&=
2
\left(
\alpha^{0_A}_{\kappa\kappa}
d^{0_B}_\kappa
d^{0_B}_\kappa
+
d^{0_A}_\kappa
d^{0_A}_\kappa
\alpha^{0_B}_{\kappa\kappa}
\right)
\quad,
\vspace{0.2cm}
\\
\tilde{C}_{3\lambda}
&=
2
\left(
\alpha^{0_A}_{\kappa\lambda}
d^{0_B}_\kappa
d^{0_B}_\lambda
+
d^{0_A}_\kappa
d^{0_A}_\lambda
\alpha^{0_B}_{\kappa\lambda}
\right)
\quad,
\vspace{0.2cm}
\\
\tilde{C}_{0\lambda}
&=
\dfrac{1}{2}
\left(
\alpha^{0_A}_{\lambda\lambda}
d^{0_B}_\lambda
d^{0_B}_\lambda
+
d^{0_A}_\lambda
d^{0_A}_\lambda
\alpha^{0_B}_{\lambda\lambda}
\right)
\quad.
\end{align}
Finally, coefficients of the distance-dependent CRP interaction potential in Eq.\eqref{eq.crp_vab_int} explicitly read
\begin{align}
V_{3\lambda}
&=
D^\prime_{3\lambda}
\left(
d^{0_A}_\lambda
+
d^{0_B}_\lambda
\right)
-
D^{\prime\prime}_{3\lambda}
\quad,
\vspace{0.2cm}
\\
V^\prime_{3\lambda}
&=
D^\prime_{3\lambda}
\left(
\alpha^{0_A}_{\lambda \lambda}
d^{0_A}_\lambda
+
\alpha^{0_B}_{\lambda \lambda}
d^{0_B}_\lambda
\right)
\quad,
\vspace{0.2cm}
\\
V_{6\lambda}
&=
D^{\prime\,2}_{3\lambda}
\quad,
\vspace{0.2cm}
\\
V^\prime_{6\lambda}
&=
D^{\prime\,2}_{3\lambda}
\left(
\alpha^{0_A}_{\lambda \lambda}
+
\alpha^{0_B}_{\lambda \lambda}
\right)
\quad.
\end{align}



\end{document}


\title{Supplementary Information:
\\
Cavity-modified electronic interactions in molecular ensembles under vibrational strong coupling: Combined insights from cavity Born-Oppenheimer perturbation and \textit{ab initio} wave function theories}

\author{Eric W. Fischer}
\email{ericwfischer.sci@posteo.de}
\affiliation{Institut f\"ur Chemie, Humboldt-Universit\"at zu Berlin, Brook-Taylor-Stra\ss{}e 2, D-12489, Berlin, Germany}

\date{\today}

\let\newpage\relax

\begin{abstract}
We provide detailed derivations for (i) the second-order CBO-PT energy corrections for dipole-dipole coupled molecular ensembles under VSC, (ii) the unimolecular CRP  and (iii) the bimolecular CRP. 
\end{abstract}

\let\newpage\relax
\maketitle
\thispagestyle{plain}

\renewcommand{\thepage}{S\arabic{page}}
\renewcommand{\theequation}{\thesection.\arabic{equation}}
\renewcommand{\thefigure}{S\arabic{figure}}
\renewcommand{\thesection}{S\arabic{section}}

\section{Derivation of CBO-PT(2) Corrections}
\setcounter{equation}{0}
We derive the CBO-PT(2) correction\cite{fischer2023} for a dipole-dipole coupled molecular ensemble under VSC. For a molecular ensemble with $M$ molecules, we consider zeroth-order product states of the form
\begin{align}
\ket{\dots\Psi^{(e)}_{0_a}\dots}
,
\ket{\dots\Psi^{(e)}_{\mu_a}\dots}
,
\ket{\dots\Psi^{(e)}_{\mu_a}\Psi^{(e)}_{\mu_b}\dots}
\,,
\end{align}
with $a,b=1,2,\dots,M$. The first state resembles the product ground state, the second one a monomer excited state and the third one a dimer excited state. 

\subsection{CBO-PT(1) Matrix Elements and Correction}
The ensemble CBO-PT(1) energy correction is determined by the matrix element
\begin{align}
E^{(1)}_\mathrm{ens}
&=
\braket{
\dots\Psi^{(e)}_{0_a}\dots
\vert
\hat{V}_1
\vert
\dots\Psi^{(e)}_{0_a}\dots}_{\underline{r}}
\quad,
\vspace{0.2cm}
\\
&=
\sum^M_a
W_a
+
\sum^M_{a,b\neq a}
\left(
\dfrac{1}{2}
V_{ab}
+
W_{ab}
\right)
\quad,
\end{align}
where we grouped two-body contributions from the dipole-dipole interaction and the DSE coupling in the second term. Matrix elements are given by
\begin{align}
W_a
&=
\braket{
\Psi^{(e)}_{0_a}
\vert
\hat{W}_a
\vert
\Psi^{(e)}_{0_a}}_{\underline{r}}
\quad,
\vspace{0.2cm}
\\
V_{ab}
&=
\braket{
\Psi^{(e)}_{0_a}
\Psi^{(e)}_{0_b}
\vert
\hat{V}_{ab}
\vert
\Psi^{(e)}_{0_a}
\Psi^{(e)}_{0_b}}_{\underline{r}}
\quad,
\vspace{0.2cm}
\\
W_{ab}
&=
\braket{
\Psi^{(e)}_{0_a}
\Psi^{(e)}_{0_b}
\vert
\hat{W}_{ab}
\vert
\Psi^{(e)}_{0_a}
\Psi^{(e)}_{0_b}}_{\underline{r}}
\quad,
\end{align}
which can be straightforwardly evaluated to
\begin{align}
E^{(1)}_a
&=
W_a
\quad,
\vspace{0.2cm}
\\
E^{(1)}_{ab}
&=
\dfrac{1}{2}
V_{ab}
+
W_{ab}
\quad,
\end{align}
since $\hat{V}_{ab}$ and $\hat{W}_{ab}$ factorize with respect to molecules $a$ and $b$. 

\subsection{CBO-PT(2) Matrix Elements}
The ensemble CBO-PT(2) correction contains two contribution
\begin{align}
E^{(2)}_\mathrm{ens}
&=
E^{(2)}_\mathrm{mo}
+
E^{(2)}_\mathrm{di}
\quad,
\end{align}
where the first term is determined by monomer excited states
\begin{align}
E^{(2)}_\mathrm{mo}
&=
\sum^M_a
\sum_{\mu_a\neq0}
\dfrac{
\vert
\braket{
\dots\Psi^{(e)}_{0_a}\dots
\vert 
\hat{V}_1
\vert 
\dots\Psi^{(e)}_{\mu_a}\dots}_{\underline{r}}
\vert^2}{\Delta^{(e)}_{\mu_a}}
\,,
\label{eq.2nd_order_ecbopt_mono}
\end{align}
and the second term includes dimer excited states
\begin{align}
E^{(2)}_\mathrm{dim}
&=
\dfrac{1}{2}
\sum^M_{a,b\neq a}
\sum_{\mu_a,\mu_b\neq0}
\dfrac{
\vert
\braket{
\dots\Psi^{(e)}_{0_a}\dots
\vert
\hat{V}_1
\vert 
\dots\Psi^{(e)}_{\mu_a}\Psi^{(e)}_{\mu_b}\dots}_{\underline{r}}
\vert^2}{\Delta^{(e)}_{\mu_a}+\Delta^{(e)}_{\mu_b}}
\,.
\label{eq.2nd_order_ecbopt_pair}
\end{align}
The matrix element in $E^{(2)}_\mathrm{mo}$ expands as
\begin{multline}
\braket{
\dots\Psi^{(e)}_{0_a}\dots
\vert 
\hat{V}_1
\vert 
\dots\Psi^{(e)}_{\mu_a}\dots}_{\underline{r}}
=
\sum^M_{a^\prime}
W_{\mu_{a^\prime}}
\delta_{aa^\prime}
\\
+
\dfrac{1}{2}
\sum^M_{a^\prime,b^\prime\neq a^\prime}
\left(
V_{\mu_{a^\prime} 0_{b^\prime}}
\delta_{aa^\prime}
+
V_{0_{a^\prime} \mu_{b^\prime}}
\delta_{ab^\prime}
\right)
\\
+
\sum^M_{a^\prime,b^\prime\neq a^\prime}
\left(
W_{\mu_{a^\prime} 0_{b^\prime}}
\delta_{aa^\prime}
+
W_{0_{a^\prime} \mu_{b^\prime}}
\delta_{ab^\prime}
\right)
\quad,
\end{multline}
which simplifies to
\begin{multline}
\braket{
\dots\Psi^{(e)}_{0_a}\dots
\vert 
\hat{V}_1
\vert 
\dots\Psi^{(e)}_{\mu_a}\dots}_{\underline{r}}
\\
=
W_{\mu_a}
+
\sum^M_{b\neq a}
\left(
V_{\mu_a 0_b}
+
2\,
W_{\mu_a 0_b}
\right)
\quad,
\end{multline}
with matrix elements
\begin{align}
W_{\mu_a}
&=
\braket{
\Psi^{(e)}_{0_a}
\vert
\hat{W}_a
\vert
\Psi^{(e)}_{\mu_a}}_{\underline{r}}
\quad,
\vspace{0.2cm}
\\
V_{\mu_a0_b}
&=
\braket{
\Psi^{(e)}_{0_a}
\Psi^{(e)}_{0_b}
\vert
\hat{V}_{ab}
\vert
\Psi^{(e)}_{\mu_a}
\Psi^{(e)}_{0_b}}_{\underline{r}}
\quad,
\vspace{0.2cm}
\\
W_{\mu_a0_b}
&=
\braket{
\Psi^{(e)}_{0_a}
\Psi^{(e)}_{0_b}
\vert
\hat{W}_{ab}
\vert
\Psi^{(e)}_{\mu_a}
\Psi^{(e)}_{0_b}}_{\underline{r}}
\quad.
\end{align}
Further, for the matrix element determining $E^{(2)}_\mathrm{di}$, one obtains
\begin{multline}
\braket{
\dots\Psi^{(e)}_{0_a}\dots
\vert
\hat{V}_1
\vert 
\dots\Psi^{(e)}_{\mu_a}\Psi^{(e)}_{\mu_b}\dots}_{\underline{r}}
\\
=
\dfrac{1}{2}
\sum^M_{a^\prime \neq b^\prime}
V_{\mu_{a^\prime}\mu_{b^\prime}}
\left(
\delta_{aa^\prime}
\delta_{bb^\prime}
+
\delta_{ab^\prime}
\delta_{a^\prime b}
\right)
\\
+
\sum^M_{a^\prime,b^\prime\neq a^\prime}
W_{\mu_{a^\prime}\mu_{b^\prime}}
\left(
\delta_{aa^\prime}
\delta_{bb^\prime}
+
\delta_{ab^\prime}
\delta_{a^\prime b}
\right)
\quad,
\end{multline}
which simplifies to
\begin{multline}
\braket{
\dots\Psi^{(e)}_{0_a}\dots
\vert
\hat{V}_1
\vert 
\dots\Psi^{(e)}_{\mu_a}\Psi^{(e)}_{\mu_b}\dots}_{\underline{r}}
\\
=
V_{\mu_a \mu_b}
+
2\,
W_{\mu_a \mu_b}
\quad,
\end{multline}
with matrix elements
\begin{align}
V_{\mu_a \mu_b}
&=
-
\sum_{\kappa,\kappa^\prime}
d^{\mu_a}_\kappa
\mathcal{T}^{\kappa\kappa^\prime}_{ab}\,
d^{\mu_b}_{\kappa^\prime}
\quad,
\vspace{0.2cm}
\\
W_{\mu_a \mu_b}
&=
\dfrac{g^2_0}{2}
\sum^{2N_c}_{\lambda,k}
d^{\mu_a}_{\lambda k}
d^{\mu_b}_{\lambda k}
\quad.
\end{align}

\subsection{CBO-PT(2) Corrections}
$E^{(2)}_\mathrm{mo}$ is determined by
\begin{multline}
\vert
\braket{
\dots\Psi^{(e)}_{0_a}\dots
\vert 
\hat{V}_1
\vert 
\dots\Psi^{(e)}_{\mu_a}\dots}_{\underline{r}}
\vert^2
\\
=
W^2_{\mu_a}
+
2\,
W_{\mu_a}
\displaystyle\sum^M_{b\neq a}
V_{\mu_a 0_b}
\\
+
\sum^M_{b,c\neq a}
V_{\mu_a 0_b}
V_{\mu_a 0_c}
+
4\,
W_{\mu_a}
\displaystyle\sum^M_{b\neq a}
W_{\mu_a 0_b}
\\
+
4
\sum^M_{b,c\neq a}
V_{\mu_a 0_b}
W_{\mu_a 0_c}
+
4
\sum^M_{b,c\neq a}
W_{\mu_a 0_b}
W_{\mu_a 0_c}
\quad,
\end{multline}
were we exploit reality of $W_{\mu_a},W_{\mu_a 0_b}$ and $V_{\mu_a 0_b}$. Terms can be grouped based on indices leading to three distinct CBO-PT(2) mono-excitation contributions
\begin{align}
E^{(2)}_a
&=
\sum_{\mu_a\neq0}
\dfrac{W^2_{\mu_a}}{\Delta^{(e)}_{\mu_a}}
\quad,
\vspace{0.2cm}
\\
\tilde{E}^{(2)}_{ab}
&=
2
\sum_{\mu_a\neq0}
\dfrac{
W_{\mu_a}
V_{\mu_a 0_b}
}{\Delta^{(e)}_{\mu_a}}
+
4
\sum_{\mu_a\neq0}
\dfrac{
W_{\mu_a}
W_{\mu_a 0_b}
}{\Delta^{(e)}_{\mu_a}}
\,,
\vspace{0.2cm}
\\
E^{(2)}_{abc}
&=
\sum_{\mu_a\neq0}
\dfrac{
V_{\mu_a 0_b}
V_{\mu_a 0_c}
}{\Delta^{(e)}_{\mu_a}}
+
4
\sum_{\mu_a\neq0}
\dfrac{
V_{\mu_a 0_b}
W_{\mu_a 0_c}
}{\Delta^{(e)}_{\mu_a}}
\vspace{0.2cm}
\\&
\hspace{2cm}
+
4
\sum_{\mu_a\neq0}
\dfrac{
W_{\mu_a 0_b}
W_{\mu_a 0_c}
}{\Delta^{(e)}_{\mu_a}}
\quad.
\nonumber
\end{align}
Note, $E^{(2)}_{abc}$ contributes to the 2-body correction for $c=b$. The pair excitation correction is determined by 
\begin{multline}
\vert
\braket{
\dots\Psi^{(e)}_{0_a}\dots
\vert
\hat{V}_1
\vert 
\dots\Psi^{(e)}_{\mu_a}\Psi^{(e)}_{\mu_b}\dots}_{\underline{r}}
\vert^2
=
V_{\mu_a \mu_b}
V_{\mu_a \mu_b}
\\
+
2\,
V_{\mu_a \mu_b}
W_{\mu_a \mu_b}
+
W_{\mu_a \mu_b}
W_{\mu_a \mu_b}
\quad,
\end{multline}
and with
\begin{align}
E^{(2)}_\mathrm{di}
&=
\dfrac{1}{2}
\sum^M_{a,b\neq a}
U^{(2)}_{ab}
\quad,
\end{align}
we find
\begin{align}
U^{(2)}_{ab}
&=
\sum_{\mu_a,\mu_b\neq0}
\dfrac{
V_{\mu_a \mu_b}
V_{\mu_a \mu_b}
}{\Delta^{(e)}_{\mu_a}+\Delta^{(e)}_{\mu_b}}
\vspace{0.2cm}
\\
&+
4
\sum_{\mu_a,\mu_b\neq0}
\dfrac{
V_{\mu_a \mu_b}
W_{\mu_a \mu_b}
}{\Delta^{(e)}_{\mu_a}+\Delta^{(e)}_{\mu_b}}
\vspace{0.2cm}
\\
&+
4
\sum_{\mu_a,\mu_b\neq0}
\dfrac{
W_{\mu_a \mu_b}
W_{\mu_a \mu_b}
}{\Delta^{(e)}_{\mu_a}+\Delta^{(e)}_{\mu_b}}
\quad.
\end{align}

\renewcommand{\thepage}{S\arabic{page}}
\renewcommand{\theequation}{\thesection.\arabic{equation}}
\renewcommand{\thefigure}{S\arabic{figure}}
\renewcommand{\thesection}{S\arabic{section}}

\section{Derivation of the Unimolecular Cavity Reaction Potential}
\setcounter{equation}{0}
We present a detailed derivation of the unimolecular CRP. We first derive the CBO-PT(2) minimizing cavity coordinate by minimizing the CBO-PT(2) energy followed by an application of the non-radiating ground state condition, which is shown to impose an additional mean-field approximation. Subsequently, we derive the unimolecular CRP at second-order of CBO-PT. Our derivation starts via the unimolecular CBO-PT(2) cPES 
\begin{align}
\mathcal{E}^{(2)}_0
&=
E^{(e)}_0
+
V_c
+
E^{(1)}_0
+
E^{(2)}_0
\quad,
\end{align}
with
\begin{align}
V_c
&=
\dfrac{\omega^2_c}{2}
x^2_c
\quad,
\vspace{0.2cm}
\\
E^{(1)}_0
&=
-
g_0\,
\omega_c\,
d^0_\lambda\,
x_c
+
\dfrac{g^2_0}{2}
d^0_\lambda
d^0_\lambda
+
\dfrac{g^2_0}{2}
\mathcal{\tilde{F}}^0_\lambda
\quad,
\vspace{0.2cm}
\\
E^{(2)}_0
&=
-
\dfrac{g^2_0}{2}
\omega^2_c\,
\alpha^0_{\lambda\lambda}
x^2_c
+
\dfrac{g^3_0}{2}
\omega_c\,
\mathcal{A}^0_{\lambda\lambda}
x_c
-
\dfrac{g^4_0}{8}
\mathcal{B}^0_{\lambda\lambda}
\quad,
\end{align}
in line with Eqs.(64)-(66) in the main text.

\subsection{Unimolecular Cavity Gradient Condition}
The CBO-PT(2) minimizing cavity coordinate, $x^{(2)}_0$, follows from the condition
\begin{align}
\dfrac{\partial}{\partial x_c}
\mathcal{E}^{(2)}_0
&=
\dfrac{\partial}{\partial x_c}
\left(
V_c
+
E^{(1)}_0
+
E^{(2)}_0
\right)
=
0
\quad,
\label{eq.cbopt2_cpes}
\end{align}
since, $\partial_{x_c}E^{(e)}_0=0$, with
\begin{align}
\dfrac{\partial}{\partial x_c}
V_c
&=
\omega^2_c\,x_c
\quad,
\label{eq.grad_vc}
\vspace{0.2cm}
\\
\dfrac{\partial}{\partial x_c}
E^{(1)}_0
&=
-
g_0\,\omega_c\,d^0_\lambda
\quad,
\label{eq.grad_e1}
\vspace{0.2cm}
\\
\dfrac{\partial}{\partial x_c}
E^{(2)}_0
&=
-
g^2_0\,
\omega^2_c\,
\alpha^0_{\lambda \lambda}\,
x_c
+
\dfrac{g^3_0}{2}
\omega_c\,
\mathcal{A}^0_{\lambda\lambda}
\quad,
\label{eq.grad_e2}
\end{align}
which leads to
\begin{align}
\omega^2_c
\left(
1
-
g^2_0\,
\alpha^0_{\lambda \lambda}
\right)
x_c
&=
g_0\,\omega_c\,d^0_\lambda
-
\dfrac{g^3_0}{2}
\omega_c\,
\mathcal{A}^0_{\lambda\lambda}
\quad.
\end{align}
From here, we find 
\begin{align}
x^{(2)}_0
&=
\dfrac{g_0\,d^0_\lambda}
{\omega_c
\left(
1
-
g^2_0\,
\alpha^0_{\lambda \lambda}
\right)}
-
\dfrac{g^3_0\,\mathcal{A}^0_{\lambda\lambda}}
{2\omega_c
\left(
1
-
g^2_0\,
\alpha^0_{\lambda \lambda}
\right)}
\quad,
\label{eq.derivative_mincavcoord}
\end{align}
and with
\begin{align}
\dfrac{1}{
1
-
g^2_0\,
\alpha^0_{\lambda \lambda}
}
&=
1
+
g^2_0\,
\alpha^0_{\lambda \lambda}
+
\mathcal{O}(g^4_0)
\quad,
\label{eq.denominator_expand}
\end{align}
we obtain
\begin{align}
x^{(2)}_0
&=
\dfrac{g_0\,d^0_\lambda}
{\omega_c
\left(
1
-
g^2_0\,
\alpha^0_{\lambda \lambda}
\right)}
-
\dfrac{g^3_0\,\mathcal{A}^0_{\lambda\lambda}}
{2\omega_c
\left(
1
-
g^2_0\,
\alpha^0_{\lambda \lambda}
\right)}
\quad,
\vspace{0.2cm}
\\
&=
\dfrac{g_0}
{\omega_c}
d^0_\lambda
+
\dfrac{g^3_0}
{\omega_c}
\alpha^0_{\lambda \lambda}
d^0_\lambda
-
\dfrac{g^3_0}
{2\omega_c}
\mathcal{A}^0_{\lambda\lambda}
+
\mathcal{O}(g^5_0)
\quad.
\label{eq.derivative_mincavcoord_final}
\end{align}
In the following, we derive $x^{(2)}_0$ from the non-radiating ground state condition, which is shown to impose the mean-field approximation, $\mathcal{A}^0_{\lambda\lambda}\approx \alpha^0_{\lambda \lambda}d^0_\lambda$, on the last term of Eq.\eqref{eq.derivative_mincavcoord_final}.

\subsection{Unimolecular Non-Radiating Ground State Condition}
In length gauge representation, the transverse electric field operator is given by\cite{schaefer2020,schnappinger2023}
\begin{align}
\underline{\hat{E}}_\perp
&=
4\pi\,
\underline{e}_\lambda
\left(
g_0\,
\omega_c\,
x_c
-
g^2_0\,
\hat{d}_\lambda
\right)
\quad,
\end{align}
such that the related CBO-PT(2) ground state expectation value reads
\begin{align}
\braket{
\Psi^{(2)}_0
\vert
\underline{\hat{E}}_\perp
\vert
\Psi^{(2)}_0}
&=
4\pi\,
\underline{e}_\lambda
\left(
g_0\,
\omega_c\,
x_c
-
g^2_0
D^{(2)}_\lambda
\right)
\quad.
\end{align}
The second-order CBO-PT dipole moment
\begin{align}
D^{(2)}_\lambda
=
\braket{\Psi^{(1)}_0\vert\hat{d}_\lambda\vert\Psi^{(1)}_0}
\quad,
\end{align}
has been derived in the SI of Ref.\cite{fischer2024} and is given by
\begin{align}
D^{(2)}_\lambda
&=
d^0_\lambda
-
\dfrac{g^2_0}{2}
\alpha^0_{\lambda\lambda}
d^0_\lambda
+
g_0\,
\omega_c\,
\alpha^0_{\lambda\lambda}\,
x_c
\quad.
\end{align}
Note, the positive sign in front of the cavity component linear in $g_0$, which results from the negative sign of the light-matter interaction term in the Pauli-Fierz Hamiltonian, which turns out to be relevant to obtain agreement between cavity gradient and non-radiating ground state conditions. Importantly, this sign change is made consistently and does not change previous results. Thus, we find for the CBO-PT(2) non-radiating ground state condition
\begin{align}
\braket{
\Psi^{(2)}_0
\vert
\underline{\hat{E}}_c
\vert
\Psi^{(2)}_0}
\propto
x_c
-
\dfrac{g_0}{\omega_c}
D^{(2)}_\lambda
=
0
\quad,
\end{align}
and explicitly that
\begin{align*}
x_c
-
\dfrac{g_0}{\omega_c}
\left(
d^0_\lambda
-
\dfrac{g^2_0}{2}
\alpha^0_{\lambda\lambda}
d^0_\lambda
+
g_0\,
\omega_c\,
\alpha^0_{\lambda\lambda}\,
x_c
\right)
&=
0
\quad,
\vspace{0.2cm}
\\
x_c
-
\dfrac{g_0}{\omega_c}
d^0_\lambda
+
\dfrac{g^3_0}{2\omega_c}
\alpha^0_{\lambda\lambda}
d^0_\lambda
-
g^2_0\,
\alpha^0_{\lambda\lambda}\,
x_c
&=
0
\quad,
\vspace{0.2cm}
\\
\left(
1
-
g^2_0\,
\alpha^0_{\lambda\lambda}
\right)
x_c
-
\dfrac{g_0}{\omega_c}
d^0_\lambda
+
\dfrac{g^3_0}{2\omega_c}
\alpha^0_{\lambda\lambda}
d^0_\lambda
&=
0
\quad,
\end{align*}
which gives
\begin{align}
x^{(2)}_0
&=
\dfrac{g_0\, d^0_\lambda}
{
\omega_c
\left(
1
-
g^2_0\,
\alpha^0_{\lambda\lambda}
\right)
}
-
\dfrac{g^3_0\, \alpha^0_{\lambda\lambda}
d^0_\lambda}
{2
\omega_c
\left(
1
-
g^2_0\,
\alpha^0_{\lambda\lambda}
\right)
}
\quad,
\vspace{0.2cm}
\\
&=
\dfrac{g_0}
{\omega_c}
d^0_\lambda
+
\dfrac{g^3_0}
{\omega_c}
\alpha^0_{\lambda \lambda}
d^0_\lambda
-
\dfrac{g^3_0}
{2\omega_c}
\alpha^0_{\lambda \lambda}
d^0_\lambda
+
\mathcal{O}(g^5_0)
\,,
\vspace{0.2cm}
\\
&=
\dfrac{g_0}
{\omega_c}
d^0_\lambda
+
\dfrac{g^3_0}
{2\omega_c}
\alpha^0_{\lambda \lambda}
d^0_\lambda
+
\mathcal{O}(g^5_0)
\quad,
\label{eq.nonrad_mincavcoord_final}
\end{align}
where we used Eq.\eqref{eq.denominator_expand} to obtain the second line. By comparison to Eq.\eqref{eq.derivative_mincavcoord_final}, we identify the CBO-PT(2) non-radiating ground state condition to impose the mean-field approximation
\begin{align}
\mathcal{A}^0_{\lambda\lambda}
=
\alpha^0_{\lambda\lambda}
d^0_\lambda
+
\tilde{\mathcal{A}}^0_{\lambda\lambda}
\approx 
\alpha^0_{\lambda\lambda}
d^0_\lambda
\,
\Leftrightarrow
\,
\tilde{\mathcal{A}}^0_{\lambda\lambda}
=
0
\quad,
\label{eq.nonrad_mincavcoord}
\end{align}
which simplifies the second and third term in Eq.\eqref{eq.derivative_mincavcoord_final} as
\begin{align}
\dfrac{g^3_0}
{\omega_c}
\alpha^0_{\lambda \lambda}
d^0_\lambda
-
\dfrac{g^3_0}
{2\omega_c}
\mathcal{A}^0_{\lambda\lambda}
\approx 
\dfrac{g^3_0}
{2\omega_c}
\alpha^0_{\lambda \lambda}
d^0_\lambda
\quad,
\end{align}
leading to the $g^3_0$-correction in Eq.\eqref{eq.nonrad_mincavcoord_final}.

\subsection{The unimolecular CRP}
\label{subsec.uni_crp}
We derive the CBO-PT(2) CRP from the corresponding cPES and minimizing cavity coordinate in Eqs.\eqref{eq.cbopt2_cpes} and \eqref{eq.nonrad_mincavcoord} with
\begin{align}
\mathcal{V}^{(2)}_\mathrm{crp}(\underline{R};\lambda)
&=
\mathcal{E}^{(2)}_0(\underline{R},x^{(2)}_0)
\quad.
\end{align}
In the following derivation, we keep only terms scaling up to $g^4_0$ in $\mathcal{V}^{(2)}_\mathrm{crp}$ in agreement with $\mathcal{E}^{(2)}_0$. We start with
\begin{multline}
\mathcal{V}^{(2)}_\mathrm{crp}
=
E^{(e)}_0
+
\dfrac{g^2_0}{2}
d^0_\lambda
d^0_\lambda
+
\dfrac{g^2_0}{2}
\mathcal{\tilde{F}}^0_\lambda
\\
+
\dfrac{\omega^2_c}{2}
\left(
1
-
g^2_0\,
\alpha^0_{\lambda\lambda}
\right)
x^{(2)}_0
x^{(2)}_0
\\
-
g_0\,
\omega_c\,
d^0_\lambda
x^{(2)}_0
+
\dfrac{g^3_0}{2}
\omega_c\,
\alpha^0_{\lambda\lambda}
d^0_\lambda
x^{(2)}_0
-
\dfrac{g^4_0}{8}
\mathcal{B}^0_{\lambda\lambda}
\quad,
\end{multline}
and obtain with Eq.\eqref{eq.nonrad_mincavcoord_final} the quadratic contribution as
\begin{multline}
x^{(2)}_0
x^{(2)}_0
=
\left(
\dfrac{g_0}
{\omega_c}
d^0_\lambda
\right)^2
+
2
\left(
\dfrac{g_0}
{\omega_c}
d^0_\lambda
\right)
\left(
\dfrac{g^3_0}
{2\omega_c}
\alpha^0_{\lambda \lambda}
d^0_\lambda
\right)
\\
+
\left(\dfrac{g^3_0}
{2\omega_c}
\alpha^0_{\lambda \lambda}
d^0_\lambda
\right)^2
\quad,
\end{multline}
which results in
\begin{align}
x^{(2)}_0
x^{(2)}_0
=
\dfrac{g^2_0}
{\omega^2_c}
d^0_\lambda 
d^0_\lambda
+
\dfrac{g^4_0}
{\omega^2_c}
d^0_{\lambda}
\alpha^0_{\lambda\lambda}
d^0_\lambda
+
\mathcal{O}(g^6_0)
\quad,
\end{align}
such that the harmonic potential term turns into
\begin{multline}
\dfrac{\omega^2_c}{2}
\left(
1
-
g^2_0\,
\alpha^0_{\lambda\lambda}
\right)
\left(
\dfrac{g^2_0}
{\omega^2_c}
d^0_\lambda 
d^0_\lambda
+
\dfrac{g^4_0}
{\omega^2_c}
d^0_{\lambda}
\alpha^0_{\lambda\lambda}
d^0_\lambda
\right)
\\
=
\dfrac{g^2_0}
{2}
d^0_\lambda 
d^0_\lambda
+
\dfrac{g^4_0}
{2}
d^0_{\lambda}
\alpha^0_{\lambda\lambda}
d^0_\lambda
-
\dfrac{g^4_0}
{2}
d^0_{\lambda}
\alpha^0_{\lambda\lambda}
d^0_\lambda
+
\mathcal{O}(g^6_0)
\quad.
\\
=
\dfrac{g^2_0}
{2}
d^0_\lambda 
d^0_\lambda
+
\mathcal{O}(g^6_0)
\quad.
\end{multline}
Further, terms linear in $x^{(2)}_0$ are given by
\begin{align}
-
g_0\,
\omega_c\,
d^0_\lambda
x^{(2)}_0
&=
-
g^2_0\,d^0_\lambda d^0_\lambda
-
\dfrac{g^4_0}
{2}
d^0_\lambda
\alpha^0_{\lambda\lambda}
d^0_\lambda
\quad,
\end{align}
and 
\begin{align}
\dfrac{g^3_0}{2}
\omega_c\,
\alpha^0_{\lambda\lambda}
d^0_\lambda
x^{(2)}_0
&=
\dfrac{g^4_0}
{2}
d^0_\lambda
\alpha^0_{\lambda\lambda}
d^0_\lambda
+
\mathcal{O}(g^6_0)
\quad.
\end{align}
Summation of terms quadratic and linear in $x^{(2)}_0$ gives
\begin{multline}
\dfrac{\omega^2_c}{2}
\left(
1
-
g^2_0\,
\alpha^0_{\lambda\lambda}
\right)
x^{(2)}_0
x^{(2)}_0
-
g_0\,
\omega_c\,
d^0_\lambda
x^{(2)}_0
+
\dfrac{g^3_0}{2}
\omega_c\,
\alpha^0_{\lambda\lambda}
d^0_\lambda
x^{(2)}_0
\\
=
-
\dfrac{g^2_0}
{2}
d^0_\lambda 
d^0_\lambda
+
\mathcal{O}(g^6_0)
\quad,
\end{multline}
which cancels the mean-field DSE component in the cPES, such that the unimolecular CRP follows as
\begin{align}
\mathcal{V}^{(2)}_\mathrm{crp}
&=
E^{(e)}_0
+
\dfrac{g^2_0}{2}
\mathcal{\tilde{F}}^0_\lambda
-
\dfrac{g^4_0}{8}
\mathcal{B}^0_{\lambda\lambda}
+
\mathcal{O}(g^6_0)
\quad.
\end{align}

\section{Derivation of the Bimolecular Cavity Reaction Potential}
\setcounter{equation}{0}
We present a detailed derivation of the bimolecular CRP. In line with the unimolecular scenario, we first discuss the bimolecular cavity gradient and non-radiating ground state conditions for CBO-PT(2), where the latter is found to consistently impose the same mean-field approximation as in the unimolecular case. Subsequently, we derive the bimolecular CRP at second-order of CBO-PT.

\subsection{Bimolecular CBO-PT(2) cPES}
The bimolecular CBO-PT(2) cPES is given by 
\begin{align}
\mathcal{E}^{(2)}_\mathrm{dimer}
&=
\tilde{\mathcal{E}}^{(2)}_A
+
\tilde{\mathcal{E}}^{(2)}_B
+
V_c
+
\mathcal{E}^{(2)}_{AB}
\quad,
\end{align}
where
\begin{align}
\tilde{\mathcal{E}}^{(2)}_X
=
E^{(e)}_{0_X}
+
E^{(1)}_{0_X}
+
E^{(2)}_{0_X}
\quad,
\quad
X=A,B
\quad,
\end{align}
as in for the unimolecular cPES and $V_c$ is a single-mode harmonic potential as before. The interaction term
\begin{align}
\mathcal{E}^{(2)}_{AB}
&=
V^{(2)}_{AB}
+
\tilde{V}^{(2)}_{AB}
+
U^{(2)}_{AB}
+
W^{(2)}_{AB}
\quad,
\end{align}
is explicitly given by Eqs.(73)-(76) in the main text.

\subsection{Bimolecular Cavity Gradient Condition}
The cavity gradient condition for the bimolecular CBO-PT(2) cPES is given by
\begin{align}
\dfrac{\partial}{\partial x_c}
\mathcal{E}^{(2)}_\mathrm{dimer}
&=
\dfrac{\partial}{\partial x_c}
\left(
\tilde{\mathcal{E}}^{(2)}_A
+
\tilde{\mathcal{E}}^{(2)}_B
+
V_c
+
\mathcal{E}^{(2)}_{AB}
\right)
=
0
\quad,
\end{align}
where the last term simplifies to
\begin{align}
\dfrac{\partial}{\partial x_c}
\mathcal{E}^{(2)}_{AB}
&=
\dfrac{\partial}{\partial x_c}
V^{(2)}_{AB}
+
\dfrac{\partial}{\partial x_c}
\tilde{V}^{(2)}_{AB}
\quad,
\end{align}
since
\begin{align}
\dfrac{\partial}{\partial x_c}
U^{(2)}_{AB}
=
\dfrac{\partial}{\partial x_c}
W^{(2)}_{AB}
=
0
\quad.
\end{align}
Cavity gradients of $\tilde{\mathcal{E}}^{(2)}_A,\tilde{\mathcal{E}}^{(2)}_B$ and $V_c$ have already been given in Eqs.\eqref{eq.grad_vc} to \eqref{eq.grad_e2}. Further, derivatives of the interaction contributions give
\begin{align}
\dfrac{\partial}{\partial x_c}
\tilde{V}^{(2)}_{AB}
&=
g^3_0\,
\omega_c
\left(
\alpha^{0_A}_{\lambda\lambda}
d^{0_B}_\lambda
+
\alpha^{0_B}_{\lambda\lambda}
d^{0_A}_\lambda
\right)
\quad,
\vspace{0.2cm}
\\
\dfrac{\partial}{\partial x_c}
V^{(2)}_{AB}
&=
-
g_0
\dfrac{D^\prime_{3\lambda}\,\omega_c}{R^3}
\quad,
\end{align}
such that the gradient condition explicitly turns into
\begin{multline}
\left(
1
-
g^2_0\,
\alpha^{0_+}_{\lambda \lambda}
\right)
x_c
=
\dfrac{g_0}{\omega_c}
d^{0_+}_\lambda
-
\dfrac{g^3_0}{2\,\omega_c}
\mathcal{A}^{0_+}_{\lambda\lambda}
\\
-
\dfrac{g^3_0}{\omega_c}
\left(
\alpha^{0_A}_{\lambda\lambda}
d^{0_B}_{\lambda}
+
\alpha^{0_B}_{\lambda\lambda}
d^{0_A}_{\lambda}
\right)
+
\dfrac{g_0}{\omega_c}
\dfrac{D^\prime_{3\lambda}}{R^3}
\quad, 
\end{multline}
with
\begin{align}
\alpha^{0_+}_{\lambda \lambda}
&=
\alpha^{0_A}_{\lambda \lambda}
+
\alpha^{0_B}_{\lambda \lambda}
\quad,
\vspace{0.2cm}
\\
d^{0_+}_\lambda
&=
d^{0_A}_\lambda
+
d^{0_B}_\lambda
\quad,
\vspace{0.2cm}
\\
\mathcal{A}^{0_+}_{\lambda\lambda}
&=
\mathcal{A}^{0_A}_{\lambda\lambda}
+
\mathcal{A}^{0_B}_{\lambda\lambda}
\quad.
\end{align}
Next, we obtain
\begin{multline}
x^{(2)}_0
=
\dfrac{g_0}{\omega_c}
\dfrac{d^{0_+}_\lambda}
{1
-
g^2_0\,
\alpha^{0_+}_{\lambda \lambda}}
-
\dfrac{g^3_0}{2\,\omega_c}
\dfrac{\mathcal{A}^{0_+}_{\lambda\lambda}}
{1
-
g^2_0\,
\alpha^{0_+}_{\lambda \lambda}}
\\
-
\dfrac{g^3_0}{\omega_c}
\dfrac{\left(
\alpha^{0_A}_{\lambda\lambda}
d^{0_B}_{\lambda}
+
\alpha^{0_B}_{\lambda\lambda}
d^{0_A}_{\lambda}
\right)}
{1
-
g^2_0\,
\alpha^{0_+}_{\lambda \lambda}}
+
\dfrac{g_0}{\omega_c}
\dfrac{D^\prime_{3\lambda}}{R^3}
\dfrac{1}
{1
-
g^2_0\,
\alpha^{0_+}_{\lambda \lambda}}
\quad,
\end{multline}
which leads with
\begin{align}
\dfrac{1}{
1
-
g^2_0\,
\alpha^{0_+}_{\lambda \lambda}
}
=
1
+
g^2_0\,
\alpha^{0_+}_{\lambda \lambda}
+
\mathcal{O}(g^4_0)
\quad,
\end{align}
to
\begin{multline}
x^{(2)}_0
=
\dfrac{g_0}{\omega_c}
d^{0_+}_\lambda
+
\dfrac{g^3_0}{\omega_c}
\alpha^{0_+}_{\lambda \lambda}
d^{0_+}_\lambda
-
\dfrac{g^3_0}{2\,\omega_c}
\mathcal{A}^{0_+}_{\lambda\lambda}
\\
-
\dfrac{g^3_0}{\omega_c}
\left(
\alpha^{0_A}_{\lambda\lambda}
d^{0_B}_{\lambda}
+
\alpha^{0_B}_{\lambda\lambda}
d^{0_A}_{\lambda}
\right)
\\
+
\dfrac{g_0}{\omega_c}
\dfrac{D^\prime_{3\lambda}}{R^3}
+
\dfrac{g^3_0}{\omega_c}
\dfrac{D^\prime_{3\lambda}\,\alpha^{0_+}_{\lambda \lambda}}{R^3}
+
\mathcal{O}(g^5_0)
\quad.
\end{multline}
After expansion of the second term in the first line as
\begin{multline}
\dfrac{g^3_0}{\omega_c}
\alpha^{0_+}_{\lambda \lambda}
d^{0_+}_\lambda
=
\dfrac{g^3_0}{\omega_c}
\left(
\alpha^{0_A}_{\lambda\lambda}
d^{0_A}_{\lambda}
+
\alpha^{0_B}_{\lambda\lambda}
d^{0_B}_{\lambda}
\right)
\\
+
\dfrac{g^3_0}{\omega_c}
\left(
\alpha^{0_A}_{\lambda\lambda}
d^{0_B}_{\lambda}
+
\alpha^{0_B}_{\lambda\lambda}
d^{0_A}_{\lambda}
\right)
\quad,
\end{multline}
we notice that the cross term in the second line of $x^{(2)}_0$ cancels such that
\begin{multline}
x^{(2)}_0
=
\dfrac{g_0}{\omega_c}
d^{0_+}_\lambda
+
\dfrac{g^3_0}{\omega_c}
\left(
\alpha^{0_A}_{\lambda\lambda}
d^{0_A}_{\lambda}
+
\alpha^{0_B}_{\lambda\lambda}
d^{0_B}_{\lambda}
\right)
-
\dfrac{g^3_0}{2\,\omega_c}
\mathcal{A}^{0_+}_{\lambda\lambda}
\\
+
\dfrac{g_0}{\omega_c}
\dfrac{D^\prime_{3\lambda}}{R^3}
+
\dfrac{g^3_0}{\omega_c}
\dfrac{D^\prime_{3\lambda}\,\alpha^{0_+}_{\lambda \lambda}}{R^3}
+
\mathcal{O}(g^5_0)
\quad.
\label{eq.derivative_mincavcoord_dimer_final}
\end{multline}
We now turn to the CBO-PT(2) non-radiating ground state condition for the interacting dimer.

\subsection{Bimolecular Non-Radiating Ground State Condition}
For the interacting dimer, the non-radiating ground state condition reads
\begin{align}
\braket{
\Psi^{(2)}_0
\vert
\underline{\hat{E}}_c
\vert
\Psi^{(2)}_0}
&\propto
x_c
-
\dfrac{g_0}{\omega_c}
\braket{\Psi^{(1)}_0\vert\hat{d}^A_\lambda+\hat{d}^B_\lambda\vert\Psi^{(1)}_0}
=
0
\quad,
\end{align}
where the dipole matrix element can be simplified as
\begin{multline}
\braket{\Psi^{(1)}_0\vert\hat{d}^A_\lambda+\hat{d}^B_\lambda\vert\Psi^{(1)}_0}
\approx
\braket{0_A\vert\hat{d}^A_\lambda\vert 0_A}
+
\braket{0_B\vert\hat{d}^B_\lambda\vert0_B}
\\
+
\braket{0_A\vert\hat{d}^A_\lambda\vert A^1_0}
+
\braket{A^1_0\vert\hat{d}^A_\lambda\vert 0_A}
\\
+
\braket{0_B\vert\hat{d}^B_\lambda\vert B^1_0}
+
\braket{B^1_0\vert\hat{d}^B_\lambda\vert0_B}
\quad,
\end{multline}
by employing orthogonality of adiabatic electronic states and neglecting higher-order perturbative terms, which are beyond linear scaling of inverse excitation energies. Further, we introduced states
\begin{align}
\ket{A^1_0}
&=
\sum_{\mu_A}
\dfrac{\braket{\mu_A\,0_B\vert\hat{W}_c\vert0_A\,0_B}}{\Delta_{\mu_A}}
\ket{\mu_A}
\quad,
\vspace{0.2cm}
\\
\ket{B^1_0}
&=
\sum_{\mu_B}
\dfrac{\braket{0_A\,\mu_B\vert\hat{W}_c\vert0_A\,0_B}}{\Delta_{\mu_B}}
\ket{\mu_B}
\quad,
\end{align}
and the bimolecular perturbation 
\begin{align}
\hat{W}_c
&=
\hat{W}_A
+
\hat{W}_B
+
\hat{W}_{AB}
+
\hat{V}_{AB}
\quad,
\end{align}
with interactions
\begin{align}
\hat{W}_{AB}
=
g^2_0\,
\hat{d}^A_\lambda
\hat{d}^B_\lambda
\quad,
\quad
\hat{V}_{AB}
=
-
\dfrac{2\hat{d}^A_\lambda\hat{d}^B_\lambda}{R^3}
\quad.
\end{align}
Evaluation of matrix elements gives
\begin{align}
\braket{0_A\vert\hat{d}^A_\lambda\vert 0_A}
&=
d^{0_A}_\lambda
\quad,
\vspace{0.2cm}
\\
\braket{0_B\vert\hat{d}^B_\lambda\vert0_B}
&=
d^{0_B}_\lambda
\quad,
\vspace{0.2cm}
\\
2
\braket{0_A\vert\hat{d}^A_\lambda\vert A^1_0}
&\approx
-
g_0\,
\omega_c\,
\alpha^{0_A}_{\lambda\lambda}\,
x_c
-
\dfrac{g^2_0}{2}
\alpha^{0_A}_{\lambda\lambda}
d^{0_A}_\lambda
\vspace{0.2cm}
\\
&\hspace{0.5cm}
-
g^2_0
\alpha^{0_A}_{\lambda\lambda}
d^{0_B}_\lambda
+
\dfrac{2}{R^3}
\alpha^{0_A}_{\kappa\lambda}
d^{0_B}_\kappa
\,,
\nonumber
\vspace{0.2cm}
\\
2
\braket{0_B\vert\hat{d}^B_\lambda\vert B^1_0}
&\approx
-
g_0\,
\omega_c\,
\alpha^{0_B}_{\lambda\lambda}\,
x_c
-
\dfrac{g^2_0}{2}
\alpha^{0_B}_{\lambda\lambda}
d^{0_B}_\lambda
\vspace{0.2cm}
\\
&\hspace{0.5cm}
-
g^2_0
\alpha^{0_B}_{\lambda\lambda}
d^{0_A}_\lambda
+
\dfrac{2}{R^3}
\alpha^{0_B}_{\kappa\lambda}
d^{0_A}_\kappa
\,,
\nonumber
\end{align}
where we exploited that matrix elements are real and symmetric, such that
\begin{multline}
\braket{\Psi^{(1)}_0\vert\hat{d}^A_\lambda+\hat{d}^B_\lambda\vert\Psi^{(1)}_0}
=
d^{0_+}_\lambda
-
g_0\,
\omega_c\,
\alpha^{0_+}_{\lambda\lambda}\,
x_c
\\
-
\dfrac{g^2_0}{2}
\left(
\alpha^{0_A}_{\lambda\lambda}
d^{0_A}_\lambda
+
\alpha^{0_B}_{\lambda\lambda}
d^{0_B}_\lambda
\right)
-
g^2_0
\left(
\alpha^{0_B}_{\lambda\lambda}
d^{0_A}_\lambda
+
\alpha^{0_A}_{\lambda\lambda}
d^{0_B}_\lambda
\right)
\\
+
\dfrac{2}{R^3}
\left(
\alpha^{0_A}_{\kappa\lambda}
d^{0_B}_\kappa
+
\alpha^{0_B}_{\kappa\lambda}
d^{0_A}_\kappa
\right)
\quad,
\end{multline}
which leads to
\begin{multline}
x^{(2)}_0
=
\dfrac{g_0}{\omega_c}
\dfrac{d^{0_+}_\lambda}{
1
-
g^2_0\,
\alpha^{0_+}_{\lambda\lambda}
}
-
\dfrac{g^3_0}{2\omega_c}
\dfrac{\left(
\alpha^{0_A}_{\lambda\lambda}
d^{0_A}_\lambda
+
\alpha^{0_B}_{\lambda\lambda}
d^{0_B}_\lambda
\right)}{
1
-
g^2_0\,
\alpha^{0_+}_{\lambda\lambda}
}
\\
-
\dfrac{g^3_0}{\omega_c}
\dfrac{
\left(
\alpha^{0_B}_{\lambda\lambda}
d^{0_A}_\lambda
+
\alpha^{0_A}_{\lambda\lambda}
d^{0_B}_\lambda
\right)
}{
1
-
g^2_0\,
\alpha^{0_+}_{\lambda\lambda}
}
+
\dfrac{g_0}{\omega_c}
\dfrac{D^\prime_{3\lambda}}
{
1
-
g^2_0\,
\alpha^{0_+}_{\lambda\lambda}
}
\dfrac{1}{R^3}
\quad.
\end{multline}
With Eq.\eqref{eq.denominator_expand}, the minimizing cavity coordinate is obtained as
\begin{multline}
x^{(2)}_0
=
\dfrac{g_0}{\omega_c}
d^{0_+}_\lambda
+
\dfrac{g^3_0}{2\omega_c}
\left(
\alpha^{0_A}_{\lambda\lambda}
d^{0_A}_{\lambda}
+
\alpha^{0_B}_{\lambda\lambda}
d^{0_B}_{\lambda}
\right)
\\
+
\dfrac{g_0}{\omega_c}
\dfrac{D^\prime_{3\lambda}}{R^3}
+
\dfrac{g^3_0}{\omega_c}
\dfrac{D^\prime_{3\lambda}\,\alpha^{0_+}_{\lambda \lambda}}{R^3}
+
\mathcal{O}(g^5_0)
\quad,
\end{multline}
which follows from the gradient equivalent in Eq.\eqref{eq.derivative_mincavcoord_dimer_final} via the mean-field approximation
\begin{align}
-
\dfrac{g^3_0}{2\,\omega_c}
\mathcal{A}^{0_+}_{\lambda\lambda}
&\approx
-
\dfrac{g^3_0}{2\,\omega_c}
\left(
\alpha^{0_A}_{\lambda\lambda}
d^{0_A}_\lambda
+
\alpha^{0_B}_{\lambda\lambda}
d^{0_B}_\lambda
\right)
\quad,
\end{align}
leading to
\begin{multline}
\dfrac{g^3_0}{\omega_c}
\left(
\alpha^{0_A}_{\lambda\lambda}
d^{0_A}_{\lambda}
+
\alpha^{0_B}_{\lambda\lambda}
d^{0_B}_{\lambda}
\right)
-
\dfrac{g^3_0}{2\,\omega_c}
\mathcal{A}^{0_+}_{\lambda\lambda}
\\
\approx 
\dfrac{g^3_0}{2\omega_c}
\left(
\alpha^{0_A}_{\lambda\lambda}
d^{0_A}_{\lambda}
+
\alpha^{0_B}_{\lambda\lambda}
d^{0_B}_{\lambda}
\right)
\quad.
\end{multline}
We finally write
\begin{align}
x^{(2)}_0
=
x^{(2)}_\mathrm{loc}
+
x^{(2)}_\mathrm{int}
\quad,
\label{eq.mincav_sep}
\end{align}
with
\begin{align}
x^{(2)}_\mathrm{loc}
&=
\dfrac{g_0}{\omega_c}
d^{0_+}_\lambda
+
\dfrac{g^3_0}{2\omega_c}
\left(
\alpha^{0_A}_{\lambda\lambda}
d^{0_A}_{\lambda}
+
\alpha^{0_B}_{\lambda\lambda}
d^{0_B}_{\lambda}
\right)
\quad,
\vspace{0.2cm}
\\
x^{(2)}_\mathrm{int}
&=
\dfrac{g_0}{\omega_c}
\dfrac{D^\prime_{3\lambda}}{R^3}
+
\dfrac{g^3_0}{\omega_c}
\dfrac{D^\prime_{3\lambda}\,\alpha^{0_+}_{\lambda \lambda}}{R^3}
\quad,
\end{align}
which separates $R$-independent/-dependent contributions.

\subsection{The Bimolecular CRP}
We write the dimer CRP as
\begin{multline}
\mathcal{V}^{(2)}_\mathrm{crp}
=
\mathcal{V}^{(2)}_\mathrm{const}
+
\dfrac{\omega^2_c}{2}
\left(
1
-
g^2_0\,
\alpha^{0_+}_{\lambda \lambda}
\right)
x^{(2)}_0
x^{(2)}_0
\\
-
g_0\,
\omega_c\,
d^{0_+}_\lambda
x^{(2)}_0
\\
+
\dfrac{g^3_0}{2}
\omega_c
\left(
\alpha^{0_A}_{\lambda \lambda}
d^{0_A}_\lambda
+
\alpha^{0_B}_{\lambda \lambda}
d^{0_B}_\lambda
\right)
x^{(2)}_0
\\
+
g^3_0\,
\omega_c
\left(
\alpha^{0_A}_{\lambda \lambda}
d^{0_B}_\lambda
+
\alpha^{0_B}_{\lambda \lambda}
d^{0_A}_\lambda
\right)
x^{(2)}_0
\\
-
g_0
\dfrac{D^\prime_{3\lambda}\,\omega_c}{R^3}
x^{(2)}_0
\quad,
\end{multline}
where all $x_c$-independent contributions are collected in $V^{(2)}_\mathrm{const}$, which explicitly given below. With Eq.\eqref{eq.mincav_sep}, we first consider the cross term of the harmonic potential
\begin{multline}
\omega^2_c
\left(
1
-
g^2_0\,
\alpha^{0_+}_{\lambda \lambda}
\right)
x^{(2)}_\mathrm{loc}
x^{(2)}_\mathrm{int}
=
g_0\,
\omega_c\,
d^{0_+}_\lambda
x^{(2)}_\mathrm{int}
\\
-
\dfrac{g^3_0}{2}
\omega_c\,
\left(
\alpha^{0_A}_{\lambda \lambda}
d^{0_A}_\lambda
+
\alpha^{0_B}_{\lambda \lambda}
d^{0_B}_\lambda
\right)
x^{(2)}_\mathrm{int}
\\
-
g^3_0\,
\omega_c\,
\left(
\alpha^{0_A}_{\lambda \lambda}
d^{0_B}_\lambda
+
\alpha^{0_B}_{\lambda \lambda}
d^{0_A}_\lambda
\right)
x^{(2)}_\mathrm{int}
+
\mathcal{O}(g^6_0)
\quad,
\end{multline}
which reduces the CRP to
\begin{multline}
\mathcal{V}^{(2)}_\mathrm{crp}
=
\mathcal{V}^{(2)}_\mathrm{const}
+
\dfrac{\omega^2_c}{2}
\left(
1
-
g^2_0\,
\alpha^{0_+}_{\lambda \lambda}
\right)
x^{(2)}_\mathrm{loc}
x^{(2)}_\mathrm{loc}
\\
+
\dfrac{\omega^2_c}{2}
\left(
1
-
g^2_0\,
\alpha^{0_+}_{\lambda \lambda}
\right)
x^{(2)}_\mathrm{int}
x^{(2)}_\mathrm{int}
\\
-
g_0\,
\omega_c\,
d^{0_+}_\lambda
x^{(2)}_\mathrm{loc}
\\
+
\dfrac{g^3_0}{2}
\omega_c
\left(
\alpha^{0_A}_{\lambda \lambda}
d^{0_A}_\lambda
+
\alpha^{0_B}_{\lambda \lambda}
d^{0_B}_\lambda
\right)
x^{(2)}_\mathrm{loc}
\\
+
g^3_0\,
\omega_c
\left(
\alpha^{0_A}_{\lambda \lambda}
d^{0_B}_\lambda
+
\alpha^{0_B}_{\lambda \lambda}
d^{0_A}_\lambda
\right)
x^{(2)}_\mathrm{loc}
\\
-
g_0
\dfrac{D^\prime_{3\lambda}\,\omega_c}{R^3}
x^{(2)}_\mathrm{int}
-
g_0
\dfrac{D^\prime_{3\lambda}\,\omega_c}{R^3}
x^{(2)}_\mathrm{loc}
\quad,
\end{multline}
\textit{i.e.}, $x^{(2)}_\mathrm{int}$ contributions in lines 3-5 have been canceled.

\subsubsection{The Dilute-Gas Limit}
We first consider the sum over $R$-independent terms in the last equation, which reads
\begin{align}
\Sigma_0
&=	
\dfrac{\omega^2_c}{2}
\left(
1
-
g^2_0\,
\alpha^{0_+}_{\lambda \lambda}
\right)
x^{(2)}_\mathrm{loc}
x^{(2)}_\mathrm{loc}
\vspace{0.2cm}
\\
&\hspace{0.5cm}
-
g_0\,
\omega_c\,
d^{0_+}_\lambda
x^{(2)}_\mathrm{loc}
\vspace{0.2cm}
\\
&\hspace{0.5cm}
+
\dfrac{g^3_0}{2}
\omega_c
\left(
\alpha^{0_A}_{\lambda \lambda}
d^{0_A}_\lambda
+
\alpha^{0_B}_{\lambda \lambda}
d^{0_B}_\lambda
\right)
x^{(2)}_\mathrm{loc}
\vspace{0.2cm}
\\
&\hspace{0.5cm}
+
g^3_0\,
\omega_c
\left(
\alpha^{0_A}_{\lambda \lambda}
d^{0_B}_\lambda
+
\alpha^{0_B}_{\lambda \lambda}
d^{0_A}_\lambda
\right)
x^{(2)}_\mathrm{loc}
\quad,
\end{align}
and reduces after making $x^{(2)}_\mathrm{loc}$ explicit followed by some simplifications to
\begin{multline}
\Sigma_0
=
-
\dfrac{g^2_0}{2}
d^{0_A}_\lambda
d^{0_A}_\lambda
-
\dfrac{g^2_0}{2}
d^{0_B}_\lambda
d^{0_B}_\lambda
-
g^2_0\,
d^{0_A}_\lambda
d^{0_B}_\lambda
\\
+
\dfrac{g^4_0}{2}
\left(
d^{0_A}_\lambda
\alpha^{0_A}_{\lambda \lambda}
d^{0_B}_\lambda
+
d^{0_A}_\lambda
\alpha^{0_B}_{\lambda \lambda}
d^{0_B}_\lambda
\right)
+
\mathcal{O}(g^6_0)
\quad.
\end{multline}
With
\begin{multline}
\mathcal{V}^{(2)}_\mathrm{const}
=
E^{(e)}_{0_A}
+
\dfrac{g^2_0}{2}
d^{0_A}_\lambda
d^{0_A}_\lambda
+
\dfrac{g^2_0}{2}
\mathcal{\tilde{F}}^{0_A}_\lambda
-
\dfrac{g^4_0}{8}
\mathcal{B}^{0_A}_{\lambda\lambda}
\\
+
E^{(e)}_{0_B}
+
\dfrac{g^2_0}{2}
d^{0_B}_\lambda
d^{0_B}_\lambda
+
\dfrac{g^2_0}{2}
\mathcal{\tilde{F}}^{0_B}_\lambda
-
\dfrac{g^4_0}{8}
\mathcal{B}^{0_B}_{\lambda\lambda}
\\
+
g^2_0\,
d^{0_A}_\lambda
d^{0_B}_\lambda
-
\dfrac{g^4_0}{2}
\left(
d^{0_A}_\lambda
\alpha^{0_A}_{\lambda \lambda}
d^{0_B}_\lambda
+
d^{0_A}_\lambda
\alpha^{0_B}_{\lambda \lambda}
d^{0_B}_\lambda
\right)
\\
-
g^4_0\,
C_0
-
g^4_0\,
\tilde{C}_0
\quad,
\end{multline}
we find that in 
\begin{align}
\mathcal{V}^{(2)}_\mathrm{crp}
&=
\mathcal{V}^{(2)}_\mathrm{const}
+
\Sigma_0
\quad,
\end{align}
the monomer mean-field DSE contributions, the first-order DSE interaction ($g^2_0$) and the $g^4_0$-interaction in the third line are canceled, such that
\begin{multline}
\mathcal{V}^{(2)}_\mathrm{crp}
=
E^{(e)}_{0_A}
+
\dfrac{g^2_0}{2}
\mathcal{\tilde{F}}^{0_A}_\lambda
-
\dfrac{g^4_0}{8}
\mathcal{B}^{0_A}_{\lambda\lambda}
\\
+
E^{(e)}_{0_B}
+
\dfrac{g^2_0}{2}
\mathcal{\tilde{F}}^{0_B}_\lambda
-
\dfrac{g^4_0}{8}
\mathcal{B}^{0_B}_{\lambda\lambda}
\\
-
g^4_0\,
C_0
-
g^4_0\,
\tilde{C}_0
\quad.
\end{multline}
which is the dimer CRP in the dilute-gas limit, $R\to\infty$.

\subsubsection{Interacting Gas}
We consider now the sum over $R$-dependent terms as
\begin{align}
\Sigma_R
&=
\dfrac{\omega^2_c}{2}
\left(
1
-
g^2_0\,
\alpha^{0_+}_{\lambda \lambda}
\right)
x^{(2)}_\mathrm{int}
x^{(2)}_\mathrm{int}
\vspace{0.2cm}
\\
&\hspace{0.5cm}
-
g_0
\dfrac{D^\prime_{3\lambda}\,\omega_c}{R^3}
x^{(2)}_\mathrm{int}
\vspace{0.2cm}
\\
&\hspace{0.5cm}
-
g_0
\dfrac{D^\prime_{3\lambda}\,\omega_c}{R^3}
x^{(2)}_\mathrm{loc}
\quad,
\end{align}
where the harmonic contribution gives
\begin{align}
\dfrac{\omega^2_c}{2}
\left(
1
-
g^2_0\,
\alpha^{0_+}_{\lambda \lambda}
\right)
x^{(2)}_\mathrm{int}
x^{(2)}_\mathrm{int}
=
\dfrac{g^4_0}{2}
\dfrac{(D^\prime_{3\lambda})^2\,\alpha^{0_+}_{\lambda \lambda}}{R^6}	
\quad.
\end{align}
The first interaction term turns into
\begin{align}
-
g_0
\dfrac{D^\prime_{3\lambda}\,\omega_c}{R^3}
x^{(2)}_\mathrm{int}
=
-
g^2_0
\dfrac{(D^\prime_{3\lambda})^2}{R^6}
-
g^4_0
\dfrac{(D^\prime_{3\lambda})^2\,\alpha^{0_+}_{\lambda \lambda}}{R^6}
\quad,
\end{align}
and the second one gives
\begin{multline}
-
g_0
\dfrac{D^\prime_{3\lambda}\,\omega_c}{R^3}
x^{(2)}_\mathrm{loc}
=
-
g^2_0
\dfrac{D^\prime_{3\lambda}\,d^{0_+}_\lambda}{R^3}
\\
-
\dfrac{g^4_0}{2}
\dfrac{D^\prime_{3\lambda}
\left(	
\alpha^{0_A}_{\lambda \lambda}d^{0_A}_\lambda
+
\alpha^{0_B}_{\lambda \lambda} d^{0_B}_\lambda
\right)}{R^3}
\quad.
\end{multline}
After summation, one obtains
\begin{multline}
\Sigma_R
=
-
g^2_0
\dfrac{D^\prime_{3\lambda}\,d^{0_+}_\lambda}{R^3}
-
g^2_0
\dfrac{(D^\prime_{3\lambda})^2}{R^6}
\\
-
\dfrac{g^4_0}{2}
\dfrac{D^\prime_{3\lambda}
\left(
\alpha^{0_A}_{\lambda \lambda}d^{0_A}_\lambda
+
\alpha^{0_B}_{\lambda \lambda}d^{0_B}_\lambda
\right)
}{R^3}
-
\dfrac{g^4_0}{2}
\dfrac{(D^\prime_{3\lambda})^2\,\alpha^{0_+}_{\lambda \lambda}}{R^6}	
\quad.
\end{multline}
From the dilute-gas limit, we know that both the $g^4_0$-vdW and the $g^4_0$-dipole-induced-dipole corrections are not cancelled during the CRP derivation. Thus, for the interacting gas, both $U_{AB}$ and $W_{AB}$ in Eqs.(75) and (76) main text contribute to the CRP. Accordingly, only the cavity-modified dipole-dipole interaction in Eq.(74) main text is altered, which leads to the interaction CRP in Eq.(83) main text
\begin{align}
\mathcal{V}^{(2)}_{AB}(R)
&=
-
\dfrac{D_3}{R^3}
+
g^2_0
\dfrac{D^{\prime\prime}_3}{R^3}
+
\Sigma_R
\vspace{0.2cm}
\\
&=
-
\dfrac{D_3}{R^3}
-
g^2_0
\left(
\dfrac{V_{3\lambda}}{R^3}
+
\dfrac{V_{6\lambda}}{R^6}
\right)
\vspace{0.2cm}
\\
&\hspace{1cm}
-
\dfrac{g^4_0}{2}
\left(
\dfrac{V^\prime_{3\lambda}}{R^3}
+
\dfrac{V^\prime_{6\lambda}}{R^6}
\right)
\quad,
\nonumber
\end{align}
with coefficients given in Eqs.(E.13)-(E.16) main text.